\documentclass[%
 reprint,
 amsmath,amssymb,
 aps,
]{revtex4-2}

\usepackage{graphicx}
\usepackage{dcolumn}
\usepackage{bm}
\usepackage{color,soul}
\usepackage{amsmath}
\usepackage{xcolor}

\begin{document}


\title{Meta-optics for spatial optical analog computing}

\author{Sajjad Abdollahramezani$^\dagger$}
\author{Omid Hemmatyar$^\dagger$}
\author{Ali Adibi$^{}$}

\email{ali.adibi@ece.gatech.edu}
\affiliation{School of Electrical and Computer Engineering, Georgia Institute of Technology, 778 Atlantic Drive NW, Atlanta, GA 30332-0250, USA}
\affiliation{$^\dagger$ Authors with equal contribution}

\date{\today}

\begin{abstract}
Rapidly growing demands for high-performance computing, powerful data processing systems, and big data necessitate the advent of novel optical devices to perform demanding computing processes effectively. Due to its unprecedented growth in the past two decades, the field of meta-optics offers a viable solution for spatially, spectrally, and/or even temporally sculpting amplitude, phase, polarization, and/or dispersion of optical wavefronts. In this Review, we discuss state-of-the-art developments as well as emerging trends in computational meta-structures as disruptive platforms for spatial optical analog computation. Two fundamental approaches based on general concepts of spatial Fourier transformation and Green’s function are discussed in detail. Moreover, numerical investigations and experimental demonstrations of computational optical surfaces and meta-structures for solving a diverse set of mathematical problems (e.g., integro-differentiation and convolution equations) necessary for on-demand applications (e.g., edge detection) are reviewed. Finally, we explore the current challenges and the potential resolutions in computational meta-optics followed by our perspective on future research directions and possible developments in this promising area.


\end{abstract}

\keywords{optical analog computing, metasurfaces, Fourier transformation, Green's function, edge sensing}

\maketitle


\section{Introduction}

For decades, optics has proved to be an unrivaled means of communication between two points as close as two chip-scale modules or as far as two intercontinental data centers. To a similar extent, optical computing has been overshadowed by powerful digital computers empowered by highly integrable transistors as mature electronic switches. However, when it comes to intricate problems such as optimization dealing with very large data sets, even breakthrough application-specific integrated circuit technologies such as tensor processing unit (TPU) have shown substantial limitations in practice \cite{jouppi2017datacenter}. All-optical computing holds the promise to solve challenging problems such as large-scale combinatorial optimizations \cite{marandi2014network,mcmahon2016fully} and probabilistic graphical models \cite{babaeian2018nonlinear}, which have found many practical applications in artificial intelligence, image processing, and social networks. 

Being an over 60-year old field, optical computing, which generally refers to the numerical computation of one- or multi-dimensional data using photons as the primary carrier medium, has been an active field of research with impressive achievements in analog/digital and classic/quantum information processing that leverage various computational approaches including Turing machines, brain-inspired neuromorphic architectures, and metaphoric systems. Optical computing as a serious rival has experienced ups and downs in different time courses \cite{ambs2010optical,athale2016optical,touch2017optical}. This stems from the indisputable superiority of optical technology in some important aspects including inherent parallel processing, low crosstalk, passive components with zero static energy consumption, and high space- and time-bandwidth products that potentially alleviate the inherent shortcomings, specifically speed, heat generation, and power hungriness, of digital electronics \cite{touch2017optical,caulfield2010future,solli2015analog}. Although early attempts have been devoted to digital optical computing by mimicking the principles of electronic computers, people found the eventual goal out of reach due to the lack of small-footprint and energy-efficient nonlinear elements working as functional optical switches \cite{tucker2010role}. In this regard, optical analog computing has gained more attracts leading to more striking developments in recent years.
Generally, analog optical computing can be classified into i) linear platforms enabling space-invariant operations (such as correlation and convolution) as well as space-variant operations (such as Hough or coordinates transform), and ii) nonlinear systems enabling pivotal operations such as logarithm transformation as well as thresholding \cite{goodman2005introduction,ambs2010optical}. Due to the vast diversity of achievements in each field, we limit the scope of this Review to the spatial optical analog computing within the linear shift-invariant (LSI) systems.

Over the past six decades, spatial analog optical computing has been arguably indebted to the progress in game-changing technologies such as diffractive and refractive optical elements, holograms, spatial light modulators (SLMs), and micro-electromechanical mirrors \cite{goodman2005introduction}. However, computing systems established based on those components are complex, challenging-to-align, and bulky is size hindering integration with compact nanophotonic circuits. Recent extraordinary advances in nanofabrication have pushed the thickness limit of various optical elements deep into the nanometric scale enabling stronger interaction of light-matter necessary for full control over the properties of incident light in miniaturized configurations. As the most interesting paradigm, meta-structures (including both metamaterials and metasurfaces \cite{engheta2006metamaterials, yu2011light, yu2014flat,  miroshnichenko2010fano, luk2010fano, atwater2011plasmonics, li2015chip, arbabi2015dielectric, kuznetsov2016optically, jahani2016all, staude2017metamaterial, genevet2017recent, hemmatyar2017phase, wang2018broadband, hemmatyar2019full, krasnok2019anomalies, reshef2019nonlinear, hail2019optical, wu2020neuromorphic, hemmatyar2020wide, arik2019beam}) hold great promise to imprint the desired transformations in amplitude, phase, and polarization of the impinging light thanks to the (sub)wavelength-scale scatterers with optimized size, shape, orientation, and composition.

\begin{figure*}[htbp]
	\centering
	\includegraphics[trim=0cm 0cm 0cm 0cm,width=.95\textwidth,clip]{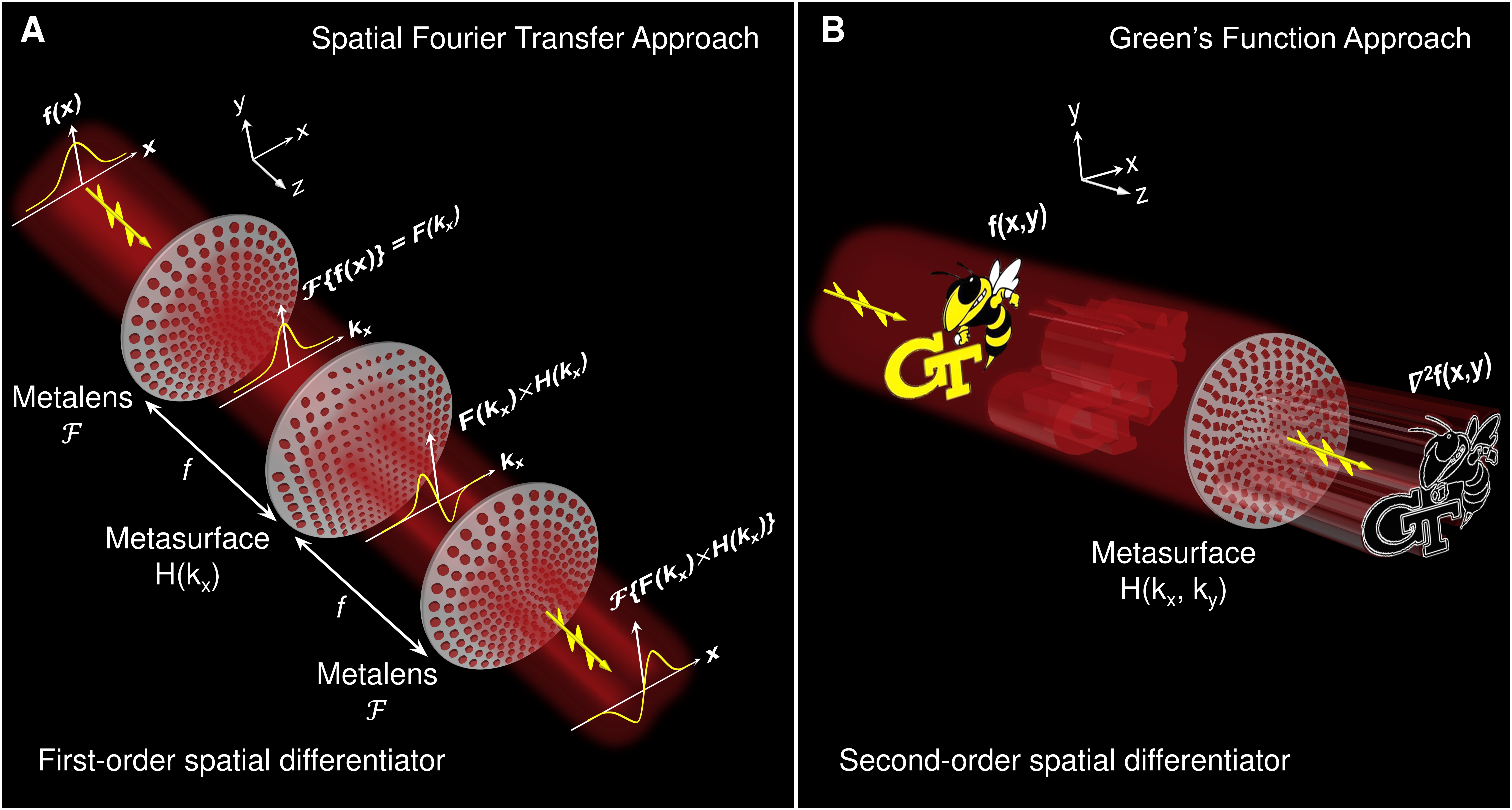}
	\caption{Computational meta-optics. Conceptual representation of spatial analog computing based on general concepts of (A) spatial Fourier transfer and (B) Green's function approach. The former leverages a 4\textit{f} correlator system formed by two series lenses with equal focal lengths and an intermediate complex-valued spatial filter located at the shared focal plane. The first metalens transforms the incident light with a spatially-variant profile (Gaussian in this example) to the Fourier domain (i.e., $k$-space) while the second metalens transfers back the light processed by the intermediate metasurface to the spatial domain. The transfer function (i.e., $H(k_{x})$) associated with the kernel of the operator of choice, first-order differentiation in this example, is encoded into the metasurface. The latter takes advantage of the nonlocal response of an engineered single-layer optical surface or a multi-layered slab that performs the desired kernel of choice. The computational system in (B) performs the second-order spatial derivation for 2D edge detection.}
	\label{figS0}
\end{figure*}

Motivated by such technological developments, Engheta and his colleagues recently introduced the concept of ``computational metamaterials'' that perform mathematical operations by direct manipulation of propagating light waves through judiciously designed metamaterial and metasurface platforms \cite{silva2014performing}. This work inspired an ever-increasing interest among several groups to propose and demonstrate viable solutions to expand the horizon of small-footprint computational systems for on-demand applications ranging from complex mathematical operations to real-time edge detection and large-scale image processing to machine vision. Two fundamental approaches have been pursued so far to realize computational meta-optics: i) spatial Fourier transfer approach, in which a 4\textit{f} system incorporating a metascreen for realization of the desired transfer function is employed, and ii) Green’s function approach, in which a resonant or non-resonant metascreen is utilized to realize the desired spatial impulse response associated with the mathematical operator of choice (see conceptual images in Figure~\ref{figS0}). In what follows, we first elaborate basic principles of each approach and then discuss relevant recent developments. These new paradigms offer real-time spatial wave-based processing mechanisms through miniaturized all-optical computing machines or potentially-integrable hardware accelerators.


\section{Spatial Fourier transfer approach}
This approach has been investigated for decades in 4\textit{f} correlators using bulky optical components \cite{goodman2005introduction}. Inspired by conventional 4\textit{f} systems, the generic architecture of an optical meta-processor is formed by the integration of a planar metamaterial or cascaded transmit/reflectarray of suitably structured metasurfaces with the Fourier transform subblocks (such as metalenses, thin lenses, or graded-index (GRIN) media). The conceptual representation of a meta-processor comprising an intermediate metasurface that realizes the desired transfer function associated with the mathematical operator of choice (here first-order spatial differentiation) and a pair of metalenses performing the exact Fourier transform of the one-dimensional (1D) field profile at their back focal planes is shown in Figure~\ref{figS0}A. Such compelling framework can significantly shrink the overall size of traditional bulky 4\textit{f} systems.

Considering a LSI system, for any given input function $f(x,y)$, the output function of the system $g(x,y)$ is calculated according to the convolution relation defined by \cite{silva2014performing}:
\begin{align} \label{equ1}
g(x,y) & =h(x,y){\ast}f(x,y) \\ \nonumber & ={\iint} h(x-x',y-y') f(x',y') \textnormal{d}x' \textnormal{d}y',
\end{align}
in which the $2$D spatial impulse response of the system is $h(x,y)$ and $\ast$ stands for the linear convolution operation. Equation~\ref{equ1} can be represented in the spatial Fourier domain as \cite{silva2014performing}:
\begin{equation} 
g(x,y)={\mathcal{F}}^{-1}\lbrace{H(k_{x},k_{y}){\mathcal{F}}\{f(x,y)\}}\rbrace,
\label{equ2}
\end{equation}
where $(k_{x}, k_{y})$ denote the two-dimensional (2D) spatial frequency variables in the Fourier space, $H(k_{x}, k_{y})$ is the spatial Fourier transform of $h(x, y)$, and $(\mathcal{F}^{-1}\{.\}) \mathcal{F}\{.\}$ represents the (inverse) Fourier transform operator. Without loss of generality, this powerful concept can be applied to a meta-processor given that $f(x, y)$ and $g(x, y)$ are the field profiles of the incident and transmitted/reflected optical beams, respectively. With this regard, the transfer function of the system, i.e., $H(k_{x}, k_{y})$, associated with the desired mathematical operator of choice can be implemented by employing a spatially-variant subblock with position-dependent transmission/reflection coefficient. It is noteworthy that the realization of an inverse Fourier transform subblock with natural materials is not practically feasible. Based on the well-known relation $\mathcal{F}\lbrace{\mathcal{F}\{A(x,y)\}}\rbrace\propto{A(-x,-y)}$, a Fourier lens can be used instead to perform inverse Fourier transform operation at the expense of image mirroring of the desired output \cite{silva2014performing}. 

\begin{figure}[htbp]
	\centering
	\includegraphics[trim=0cm 0cm 0cm 0cm,width=0.5\textwidth,clip]{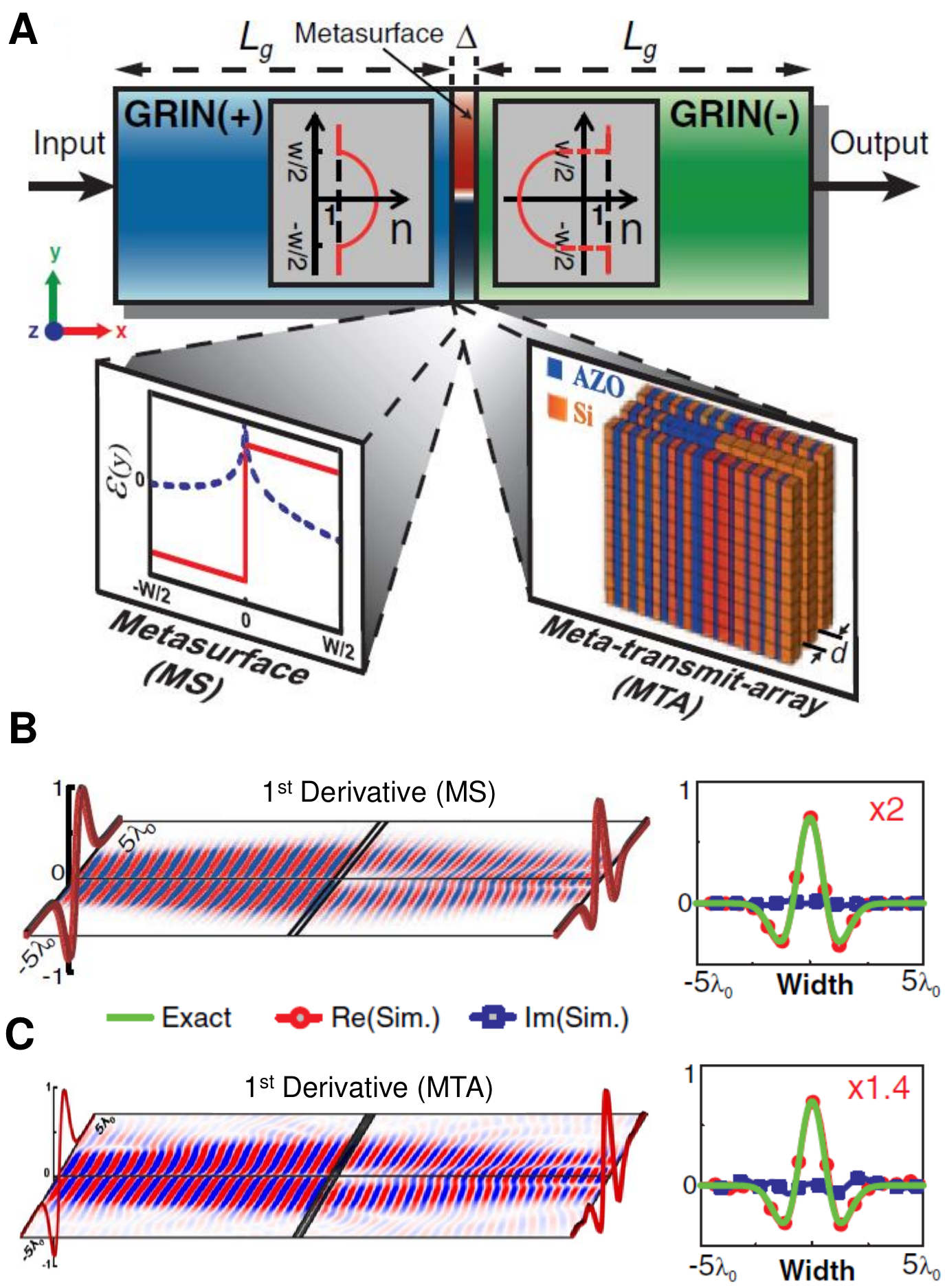}
	\caption{Metamaterials for spatial analog computing. \cite{silva2014performing}.
	(A) Cascaded properly designed GRIN(+)/meta-structure/GRIN(-) system to perform mathematical operations in the spatial Fourier domain. Two designs are proposed for the middle meta-structure including (i) an specific-purpose thin ($\Delta = \lambda_{0}/3$) single-layer metasurface with the prescribed permitivity and permeability (left inset), and (ii) a general-purpose transmitarray of metasurfaces (so-called MTA) comprised of three metasurfaces formed by an array of hybrid plasmonic-dielectric nanocubes that provide a rich set of transmission amplitude and phase profiles.
	A snapshot on the evolution of the z-component of the electric field  through the system operating as the first-order differentiator formed based on (B) the first scenario (i.e., i) and (C) second scenario (i.e., ii). Simulation results (both real and imaginary parts) are compared with the analytical solutions in the insets.
	}
	\label{figS1}
\end{figure}

Many scientific phenomena and engineering problems such as heat transfer, electromagnetic radiation, Kirchhoff's second law in circuit analysis can be described through a system of partial differential, integral, and integro-differential equations. As such, finding the exact solution to a given set has been the main focus of many works in the field of analog computing. To this end, we centralized the rest of discussion around the solution of these types of problems. Based on Fourier transform principles, the $n^\textrm{th}$ derivative of a 2D wavefunction, to which an arbitrary function is decomposed, is linked to the corresponding wavenumber and the first spatial Fourier transform according to \cite{goodman2005introduction}:
\begin{equation} 
\nabla^{n}f(x,y)={{\mathcal{F}}^{-1}}\lbrace(ik)^{n}{\mathcal{F}}\{f(x,y)\}\rbrace
\label{equ5}
\end{equation}
in which $i=\sqrt{-1}$ and $k = \sqrt{k_{x}^2+k_{y}^2}$. Given the continuum of wavenumbers forming the basis of field expansion, the transfer function corresponding to the $n^\textrm{th}$ derivation follows a parabola profile with the order of $n$. Without loss of generality, we limit the problem to more practical optical computing problems dealing with one transverse variable. In this regard, we assume the continuum of wavenumbers is called $x$ which represents wavenumber, and not the spatial variable, along the $x$-axis. Thus the desirable transfer function is in the form of $H(x) \propto (ix)^{n}$. Since in the passive (gain-less) media the transmission coefficient cannot surpass unity, the transfer function has to be normalized to the lateral size of the structure $D$, i.e., $H(x) \propto (ix/(D/2))^{n}$. When it comes to the integration operation, the required transfer function is defined as $H(x) \propto (d/ix)$ in the spatial Fourier domain with a singularity at $x = 0$, which has to be handled. It has been proven that imposing constant value of unity on $H(x)$ in the small region $d$ (typically an order of magnitude less than $D$) near the origin well mitigates any need for gain requirements \cite{silva2014performing}. Upon calculation of the transfer function of the desired operator of choice, the rest is dealing with the proper discretization in finite steps of the phase and amplitude of the transfer function and assigning the appropriate elements of the metasurface to encode the transmission or reflection of the incident field. In what follows, different frameworks for hardware implementation of the desired transfer function are discussed in detail.

In a seminal work, Silva $et~al.$ proposed to use a system comprised of three cascaded subblocks of Fourier transform, an inhomogeneous spatial Fourier filter, and inverse Fourier transform \cite{silva2014performing}. As shown in Figure~\ref{figS1}A, for the Fourier transform (GRIN(+)) and inverse Fourier transform (GRIN(-)) subblocks, 2D GRIN media characterized by length $L_{g}$, unit relative permeability (i.e., $\mu_{r} = \pm 1$), and parabolic-shaped permittivity, defined as $\epsilon_{y} =\pm \epsilon_{c}[1-(\pi/2L_{g})^2y^2]$, with positive and negative sign for Fourier and inverse Fourier transform, respectively, are considered. For the intermediate subblock (i.e., spatial Fourier filter), either a planar thin metamaterial with transversally inhomogeneous optical properties $\epsilon_\textrm{mm}(y)/\epsilon_{0} = \mu_\textrm{mm}(y)/\mu_{0}$ or a transmitarray of metasurfaces, which consists of composite plasmonic-dielectric nanobricks \cite{monticone2013full}, can be used. The transmitarray of metasurfaces is wisely engineered to encode the desired phase and amplitude patterns into the transmitted wave while minimizing disturbing reflections, which is necessary for high-efficient implementation of any desired local transfer function. Different mathematical operations were explored including first- and second-order differentiation, integration, and convolution. For the sake of brevity, we limit the discussion to the first-order spatial derivation.

Considering the thickness of the planar metamaterial as $\Delta<\lambda_{0}$, where $\lambda_{0}$ is the free-space wavelength of the infrared light, the relative permittivity and permeability associated with the normalized transfer function, i.e., $H(y)\propto{iy/(W/2)}$, is described as \cite{silva2014performing}:
\begin{equation} 
\frac{\epsilon_\textrm{mm}(y)}{\epsilon_{0}} = \frac{\mu_\textrm{mm}(y)}{\mu_{0}} = i(\frac{\lambda_{0}}{2\pi\Delta})\textrm{ln}(\frac{-iW}{2y}).
\label{equ3}
\end{equation}	
Figure~\ref{figS1}B illustrates the field distribution evolution (snapshot in time) as the wave propagates through the computational system. Comparison of the simulated electric field distribution at the output plane ($2L_{g} + \Delta$) and the first-order spatial derivative calculated analytically, given that the input function is $f(y) \propto \textrm{exp}(-y^{2})$, corroborates the performance of this approach. As a more practically feasible approach, the authors leveraged a transmitarray of metasurfaces cascaded with two GRIN (+) media to implement the first-order spatial derivation. Figure~\ref{figS1}C represents the comparison assessment between numerical simulations and analytical results. 

\begin{figure}[htbp]
	\centering
	\includegraphics[trim=0cm 0cm 0cm 0cm,width=0.5\textwidth,clip]{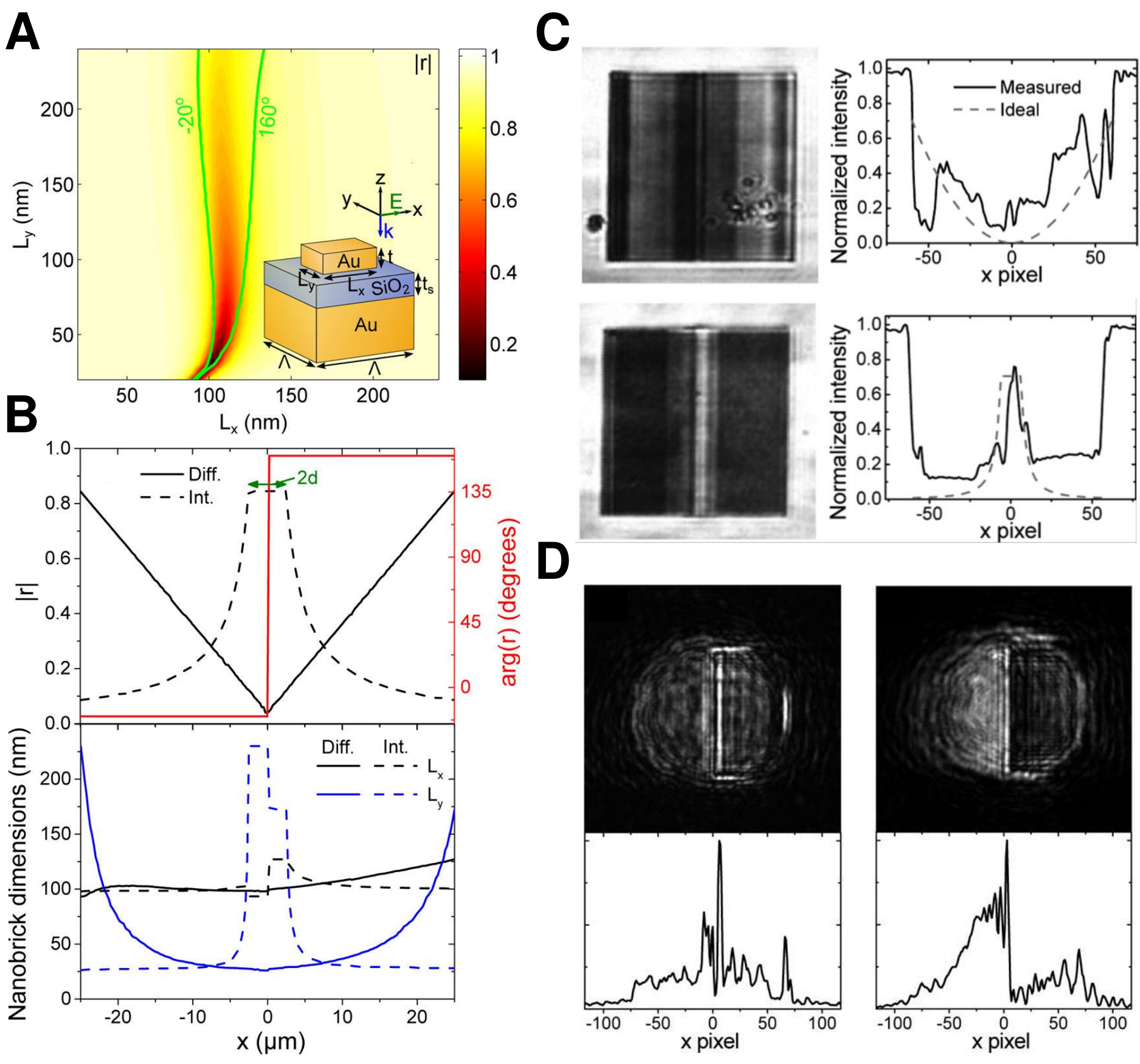}
	\caption{Reflective plasmonic metasurfaces for spatial analog signal processing \cite{pors2015analog}.
	(A) Color map of the calculated reflection coefficient as a function of the width of nanobricks for a reflective metasurface comprised of an array of Au nanobricks on top of an Au substrate separated with a thin oxide layer. Green curves indicate two contours of reflection phase with $\pi$-phase difference.
	(B) Position-dependent reflection amplitude/phase profile (top panel) and the corresponding nanobrick width (bottom panel) for a 50 $\mu$m-wide differentiator and integrator.
	(C) Bright-field images and the measured average normalized reflectivity from the metasurface realizing the transfer function of differentiation (top panel) and integration (bottom panel).
	(D) Reflected far-field intensity (top panel) and intensity profile averaged along the central part in the y-direction as a function of the x-coordinate (bottom panel) for the differentiator (left panel) and the integrator (right panel) system. Here, the excitation is through a bulky optical lens acting as Fourier and inverse Fourier subblocks simultaneously. 
	}
	\label{figS2}
\end{figure}

\begin{figure}[htbp]
	\centering
	\includegraphics[trim=0cm 0cm 0cm 0cm,width=0.5\textwidth,clip]{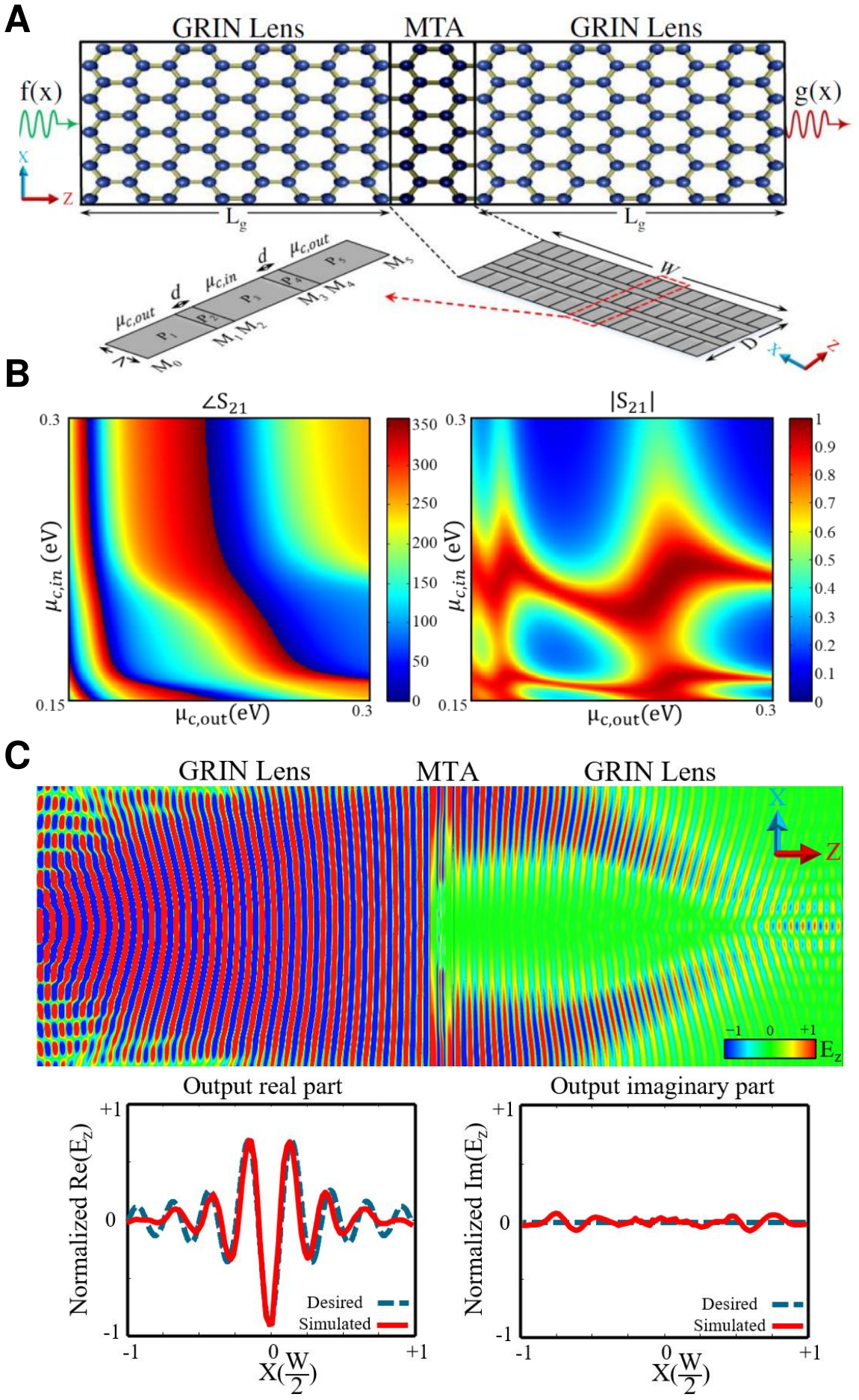}
	\caption{Linear transversely invariant graphene-based meta-structure to perform mathematical operations \cite{abdollahramezani2015analog}.
	(A) Sketch of a 2D graphene-based computing system consisting of two cascaded GRIN lenses incorporating a transmitarray of three symmetric stacked metalines with total length of $D$ and width of $W$ (inset: basic unit cell of the metaline).
	(B) Transmission phase and amplitude of the metaline as a function of the chemical potential associated with the surface conductivity.
	(C) Snapshots of the z-component of the electric field distribution through the GRIN lens/transmitarray of metalines/GRIN lens system as a second-order differentiator given that the input wave is a Sinc function.
	}
	\label{figS3}
\end{figure}

To achieve strongly independent phase and amplitude modulation, Pors and co-workers utilized the rich nature of gap surface plasmon mode in a plasmonic meta-reflectarray \cite{pors2015analog}. As shown in Figure~ \ref{figS2}A, their proposed structure consists of an array of gold (Au) nanobricks on a metal substrate separated by an oxide spacer layer. Such a confined mode excited in the gap between the top metallic nanoresonator and the substrate propagates back and forth between the two ends of the nanoresonator leading to the energy leakage to free space. Strong modulation granted by such a Fabry-Perot-like resonance within the engineered inclusion facilitates manipulation of the scattered near-infrared light both spectrally and spatially. As shown in Figures~\ref{figS2}B and \ref{figS2}C, it is evident that the fabricated computational metasurfaces fairly follow the features of their designed transfer functions while discrepancies exist in the spatial variation of the reflectance due to mainly fabrication tolerances, uncertainty in the material optical properties, and interference of high-spatial frequencies reflected from the bare substrate. Hence, the proof-of-concept demonstration of the differentiator and integrator are not in a very good agreement with the numerical calculations as depicted in Figure~\ref{figS2}D. However, this work is the first experimental implementation in the field of computational meta-optics.

\begin{figure}[htbp]
	\centering
	\includegraphics[trim=0cm 0cm 0cm 0cm,width=0.5\textwidth,clip]{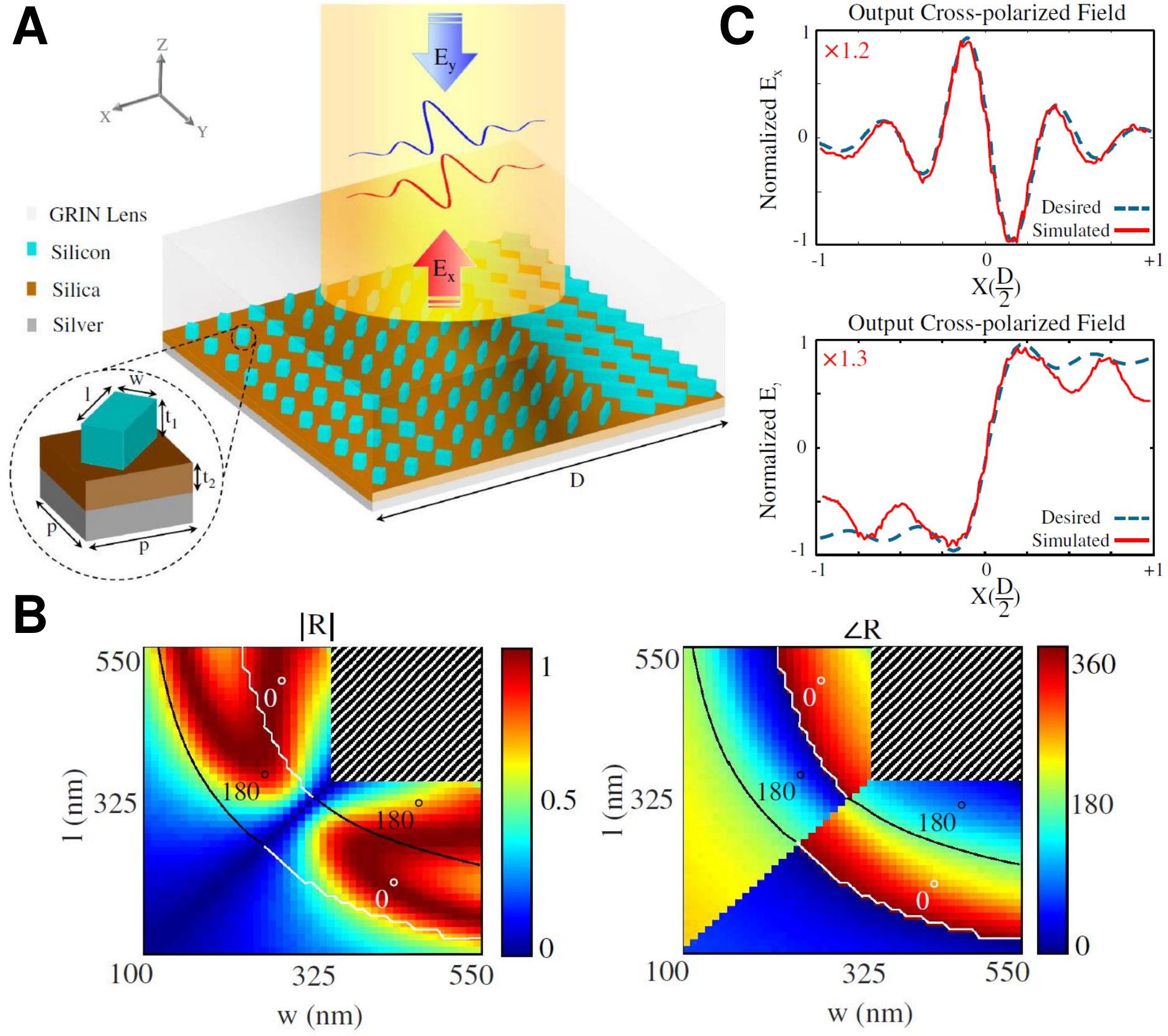}
	\caption{Analog optical computing using hybrid dielectric-plasmonic metasurfaces \cite{chizari2016analog}. 
	(A) Schematic representation of the dielectric GRIN media/hybrid metasurface to perform mathematical operations including spatial differentiation and integration. The metasurface is composed of an arrangement of Si nanobricks on an oxide layer deposited on an Ag substrate.
	(B) Simulated amplitude and corresponding phase profiles of the reflected cross-polarized light considering the rotation angle of 45$^{\circ}$ with respect to the x-axis for the constituent nanoresonator.
	(C) Comparison of the simulated reflected cross-polarized field and the exact analytical solution for the first-order differentiator (top panel) and first-order integrator (bottom panel). 
	}
	\label{figS4}
\end{figure}

Shrinking the conventional bulky 4\textit{f} correlator to on-chip integrable configurations are highly desirable for next-generation highly miniaturized computational circuits. Motivated by that, Abdollahramezani $et~al.$ recently introduced the concept of ``metalines'' that endows full control over the amplitude and phase profiles of the propagating graphene plasmons in a dynamic fashion \cite{abdollahramezani2015analog}. As shown in Figure~\ref{figS3}A, a transmitarray of such metalines, whose surface conductivity can be locally and independently controlled through height encoding of the substrate beneath the graphene layer (or equivalently an external bias), enables realization of any transfer function in an ultrathin, integrable, and truly planar platform. Due to the high confinement of graphene plasmons that empowers the integration of the Fourier transform subblock in a cascaded structure, the presented 2D configuration is orders of magnitude smaller than the traditional metasurface-based counterparts. To facilitate the calculation of surface conductivity-dependent transmission/reflection coefficients (Figure~\ref{figS3}B) of the piecewise-constant elements of the dynamic metaline, the authors leveraged a representative simplified analytical treatment on the scattering of graphene plasmons from inhomogeneous lateral hetrostructures. Figure~\ref{figS3}C represents a comparison assessment of the computational system (i.e., GRIN lens/transmitarray of metalines/ GRIN lens) designed to perform the second order spatial derivation upon excitation with a beam with Sinc profile.

\begin{figure}[t]
	\centering
	\includegraphics[trim=0cm 0cm 0cm 0cm,width=0.5\textwidth,clip]{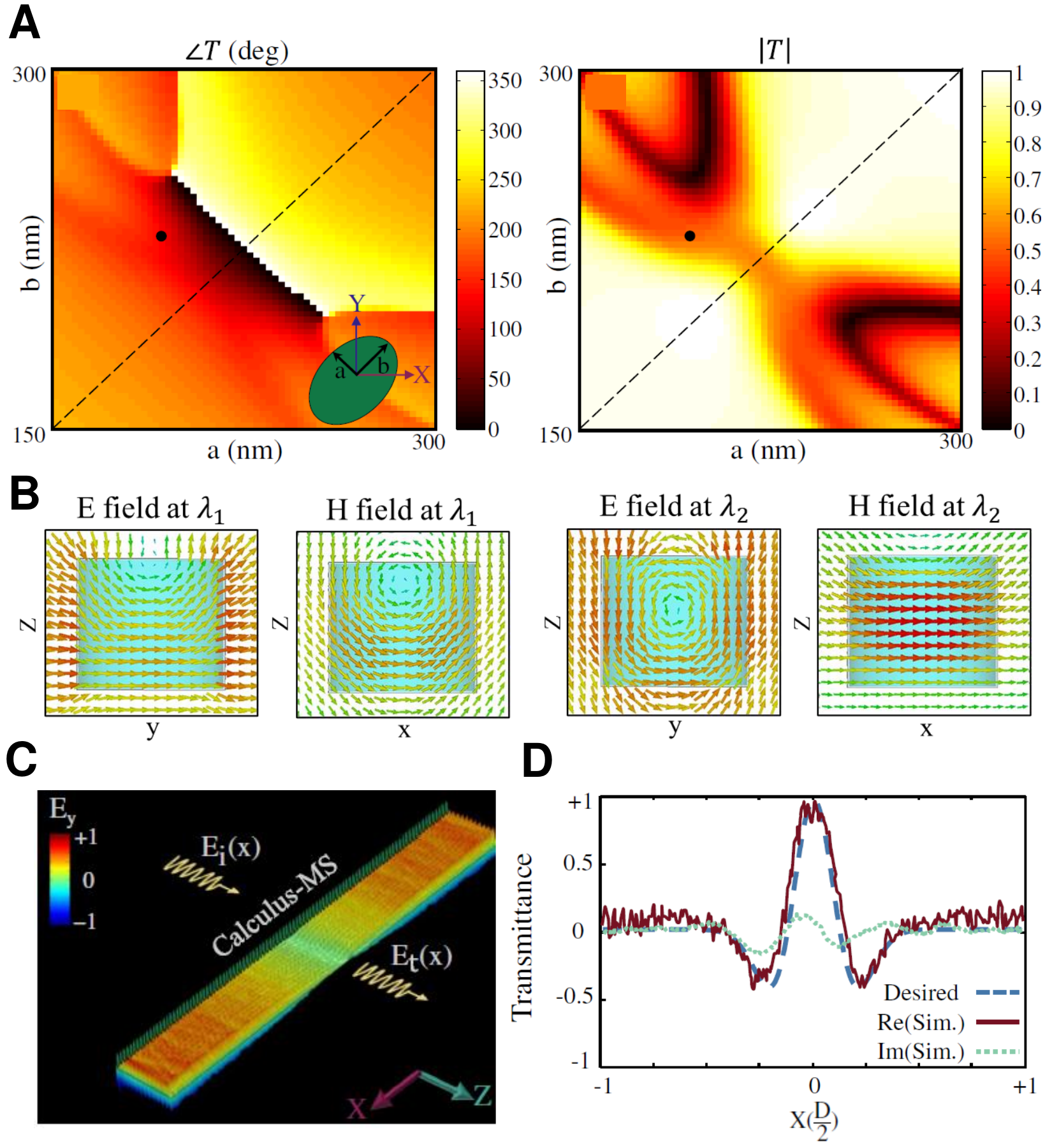}
	\caption{All-dielectric metasystems as equation solvers \cite{abdollahramezani2017dielectric}.
	(A) Numerically calculated phase and amplitude of transmitted light from the computational metasurface as a function of minor and major axis dimension.
	(B) Simulated electric and magnetic lines in the cross section of a Si nanodisk at the electric and magnetic resonances fulfilled at structural parameters associated with the dots shown in (A).
	(C) Snapshot of the y-component of the simulated electric field propagating through the computational metasurface realizing the kernel for solving the integro-differentiation equation.
	(D) Comparison assessment between the normalized electric field at the output of the proposed metalens/computational metasurface/metalens system and analytical results for integro-differential equation solving.
	}
	\label{figS5}
\end{figure}

\begin{figure}[htbp]
	\centering
	\includegraphics[trim=0cm 0cm 0cm 0cm,width=0.5\textwidth,clip]{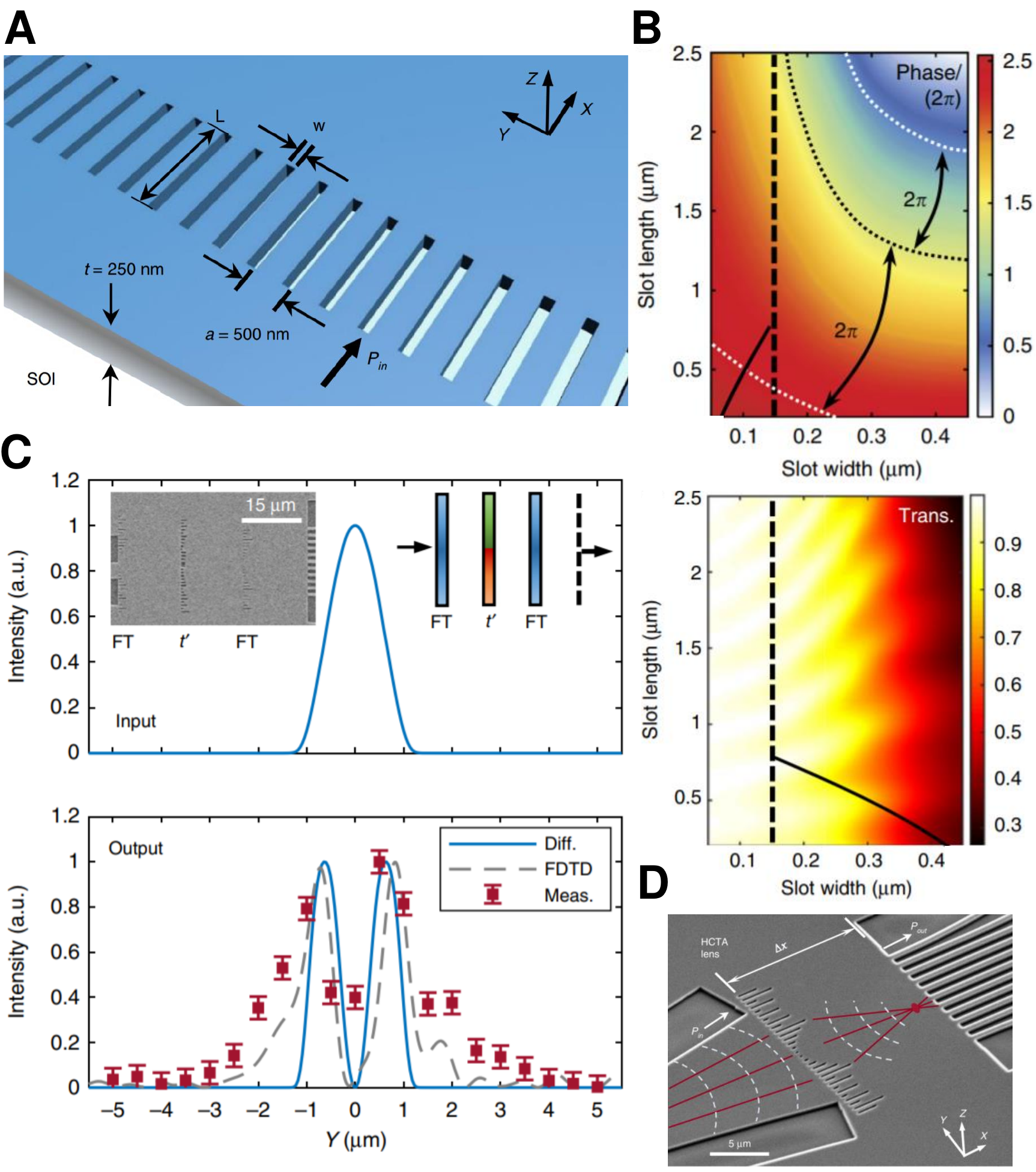}
	\caption{On-chip analog computing based on a SOI platform \cite{wang2019chip}.
	(A) Schematic representation of an on-chip high contrast tranmitarray defined on a SOI substrate to manipulate the transmittance and phase response of the in-plane propagating wave.
	(B) Simulated amplitude (top panel) and phase (bottom panel) spectra of the transmitted field as functions of the slot width (w) and height (h) for an incident wavelength of 1550 nm. Contours with 2$\pi$ phase-shift are depicted by dotted lines.
	(C) The input incident wave profile (top panel) and analytical, simulated, and measured distribution at the output (bottom panel) of the on-chip differentiator. The conceptual design and SEM image of the fabricated device are shown in the inset.
	(D) SEM image of the fabricated on-chip metalens located in front of a mode expander and coupled to eleven single mode waveguide to facilitate characterization of the spatial distribution of focused light.
	}
	\label{figS6}
\end{figure}

To harness the full potential of plasmonic and dielectric modes within a single structure, Chizari $et~al.$ proposed to use a metasurface comprised of an engineered array of size-variant nanoresonators on top of a silver (Ag) backreflector separated with an oxide intermediate layer (see Figure~\ref{figS4}A) \cite{chizari2016analog}. The combination of low-aspect ratio rotating silicon (Si) nanobricks, which support the low-loss Mie resonance, and the lossy back-reflector, which facilitates multiple reflections within the low-index spacer layer, grants unique agility on the required phase delay and amplitude modulation of the cross-polarized light. As shown in Figure~\ref{figS4}B, the designed unit cell can impart full $2\pi$ phase coverage corresponding with the desired amplitude in the range of (0, 1) on the reflected near-infrared light. For the full operation, such a compact design only needs one coupled lens serving as both Fourier and inverse Fourier transform subblocks simultaneously. Figure~\ref{figS4}C shows that a good agreement is achieved between the numerical simulation results and the designed first-order differentiator and integrator using such a CMOS-compatible metasurface.


While the reflective plasmonic metasurfaces generally provide larger phase span thanks to the added Fabry-Perot channel, transmittive metasurfaces grant easy access to the processed field profile at their output, which is a must for cascadable optical systems. However, optical metasurfaces supporting plasmonic resonances exhibit intrinsic nonradiative losses and limited scattering cross sections. To mitigate these challenges, high-contrast metasurfaces have recently garnered significant attentions in the field of nanophotonics. Abdollahramezani and co-workers recently demonstrated that an all-dielectric metasurface based on array of anisotropic Si nanoresonators can fully tailor the transmission amplitude and phase responses locally and almost independently (see Figure~\ref{figS5}A), which can facilitate the realization of mathematical operators \cite{abdollahramezani2017dielectric}. Such a unique feature is granted by the magnetic dipole moment on account of the circular displacement current excitation and the electric dipole moment due to considerable charge accumulation at the corner of each resonator (see Figure~\ref{figS5}B). Thanks to the the interplay between the first and second Mie resonant modes (i.e., magnetic and electric dipole moment) achieved by changing the major-to-minor axis aspect ratio of the nanoresonator, fairly any desired transformation of the amplitude and phase can be imparted on the transmitted light (see Figure~\ref{figS5}A). The authors leveraged the full potential of a cascaded, compact platform of metalens/computational metasurface/metalens to perform all-optical signal processing including constant coefficient integro-differential equation (see Figures~\ref{figS5}C and \ref{figS5}D), which has not been shown previously. Taking advantage of the incorporated metalenses as the functional Fourier and inverse Fourier transform subblocks, rather than the bulky conventional lenses, paves the way for highly integrable on-demand computational systems.

Considering well-established foundry-based silicon photonics, implementation of on-chip mathematical operators promises the next generation of small footprint, low-power consumption, and multi-purpose computational photonic integrated circuits. More recently, Wang and co-workers demonstrated parallel signal processing by leveraging a 1D on-chip high-contrast transmitarray of metasurfaces (see Figure~\ref{figS6}A) \cite{wang2019chip}. By judiciously adjusting the width and length of the void slots in the Si-on-insulator (SOI) substrate, complete control of the transmitted amplitude and phase profile can be achieved over a high bandwidth (see Figure~\ref{figS6}B). Figure~\ref{figS6}C illustrates experimental results of the fabricated three-layer system necessary for successful implementation of on-chip mathematical operators. While the first and second metalenses perform the Fourier and inverse Fourier transform, the spatially varying transmission coefficient is encoded into the middle mask layer to perform the required 1D transfer function (see inset in Figure~\ref{figS6}C). The calculated analytical, numerically simulated, and measured spatial spectra are in good alignment for the spatial differentiator as shown in the bottom panel of Figure~\ref{figS6}C. Exploiting an integrated metalens (shown in Figure~\ref{figS6}D) significantly miniaturizes the overall size of the on-chip photonic processor.

\begin{figure}[htbp]
	\centering
	\includegraphics[trim=0cm 0cm 0cm 0cm,width=0.5\textwidth,clip]{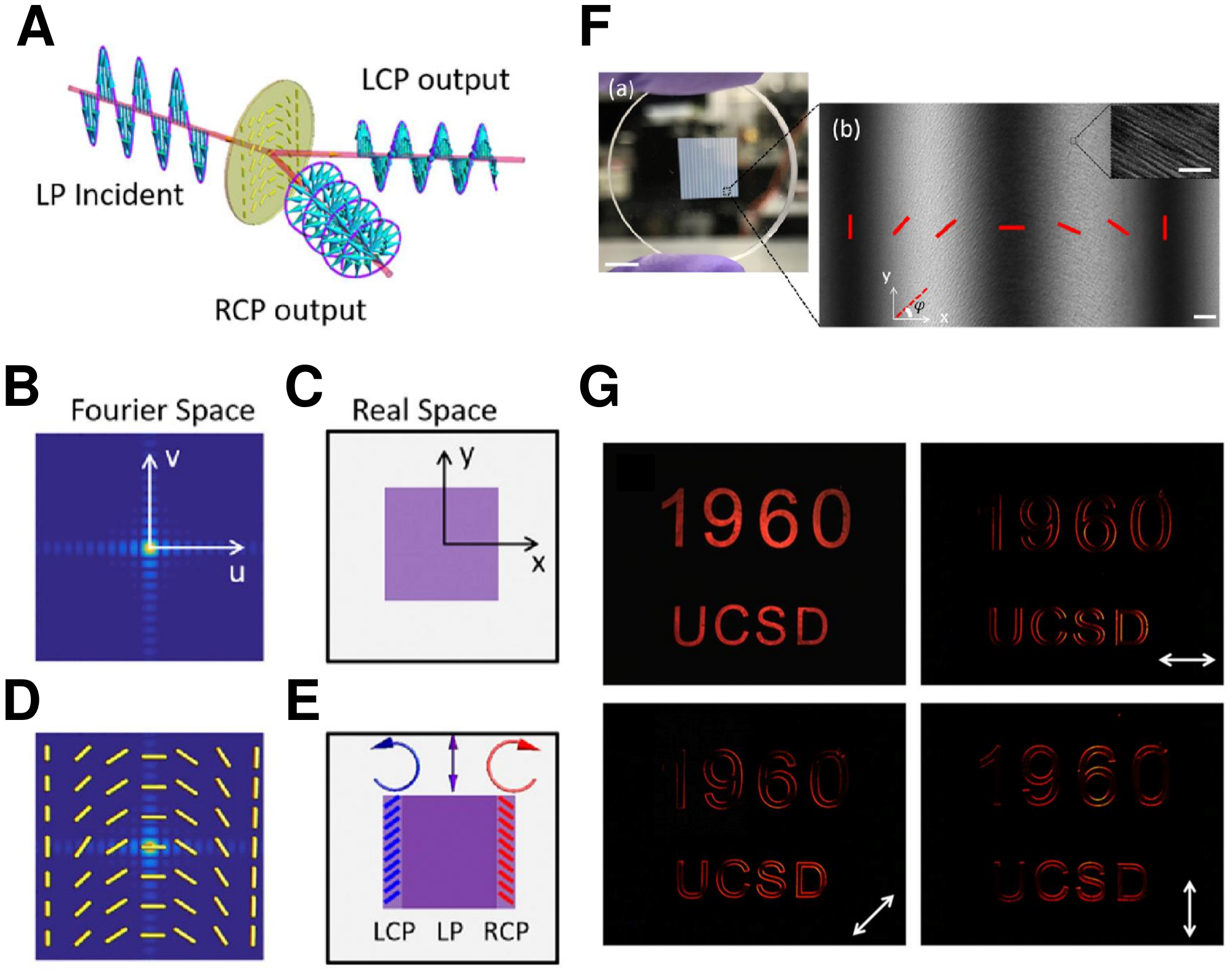}
	\caption{Edge detection using PB-phase gradient metasurfaces \cite{zhou2019optical}. (A) The concept of PB-phase. Upon illumination of an incident LP beam, the PB-phase metasurface splits the light into LCP and RCP beams with opposite directions. The Fourier space spectrum (B) and real-space image (C) of a square object. When a PB-phase gradient metasurface is added at the Fourier plane, (B) and (C) change to (D) and (E), respectively. (F) The fabricated metasurface with a pattern area of 8 mm $\times$ 8 mm embedded in a 3 mm-thick glass substrate (scale bar is 5 mm). The polariscope optical image of the area marked in the left panel is shown in the right panel (scale bar is 25 $\mu$m). The constituent elements in one period of the metasurface are indicated by the red bars in the right panel, and the inset shows the SEM micrograph of the fabricated device (scale bar is 500 nm). (G) The top-left panel shows the original image, while other three panels show detected images at the output of the system. Different orientation of the intermediate metasurface, which is indicated by the white arrows, resolves different portion of edges.}
	\label{figO8}
\end{figure}

More recently, Zhou \textit{et al.} have shown experimental realization of first-order spatial differentiation and demonstrated broadband 1D edge detection by leveraging the spin-orbit interaction of light and a properly designed Pancharatnam-Berry(PB)-phase metasurface sandwiched between two orthogonally aligned laser-written linear Glan polarizers \cite{zhou2019optical}. The PB-phase gradient metasurface splits the incident linearly polarized (LP) beam into left-handed and right-handed circularly polarized (LCP and RCP) beams propagating in opposite directions (see Figure~\ref{figO8}A). To explore the effectiveness of the proposed approach, the edges of an squared-shape object (slit) are detected in an analog fashion. The real-space image (i.e., the electric field distribution of the object in the spatial domain) and the Fourier space spectrum of the object are shown in Figures~\ref{figO8}C and ~\ref{figO8}B, respectively. By introducing a PB-phase gradient metasurface at the Fourier plane (see Figures~\ref{figO8}D), the LCP and RCP components of the output electrical field at the image plane obtain opposite phase gradients. This, in turn, results in a slight shift of LCP and RCP images in opposite directions as shown in Figure~\ref{figO8}E. When this output image passes through the analyzer (i.e., the orthogonal Glan polarizer along y-direction implemented right after the metasurface), only two shaded areas are left at the output that indicate the edges of the input image. The fabricated metasurface with 8 mm $\times$ 8 mm pattern area embedded inside a 3 mm-thick glass substrate is shown in Figures~\ref{figO8}F. The metasurface pattern is written by a femtosecond pulsed laser beam focused 50 $\mu$m beneath the surface of the glass. Interestingly, this edge detection technique enables tunable resolution at the resultant edges by varying the PB-phase gradient period. Moreover, as shown in Figure~\ref{figO8}G, this edge detection technique is sensitive to the orientation of the phase gradient metasurface (i.e., the gradient direction of the 1D metasurface indicated by white arrows). The use of thick glass-based meta-structure in their implementation makes the metasurface highly efficient in term of transmitted power (around 90\%) which is the main advantage over the plasmonic metasurfaces. Additionally, since the operational bandwidth of this approach is not limited by the critical plasmonic coupling condition \cite{zhu2017plasmonic}, realization of broadband transfer functions is guaranteed. An interesting extend of this work can be implementation of a broadband highly efficient 2D edge detector.

In addition to so far demonstrated platforms enabling mathematical operators, several other theoretical works exist in the literature using spatial Fourier transformation concept to realize first-order differentiation using bilayered metasurfaces \cite{farmahini2013metasurfaces}, differential and integral operations using Ag dendritic metasurfaces \cite{chen2017performing,chen2020quasi}, and multi-way parallel mathematical operations based on discrete metamaterials \cite{wu2018arbitrary}.

\section{Green's function approach}

\begin{figure}[htbp]
	\centering
	\includegraphics[trim=0cm 0cm 0cm 0cm,width=0.5\textwidth,clip]{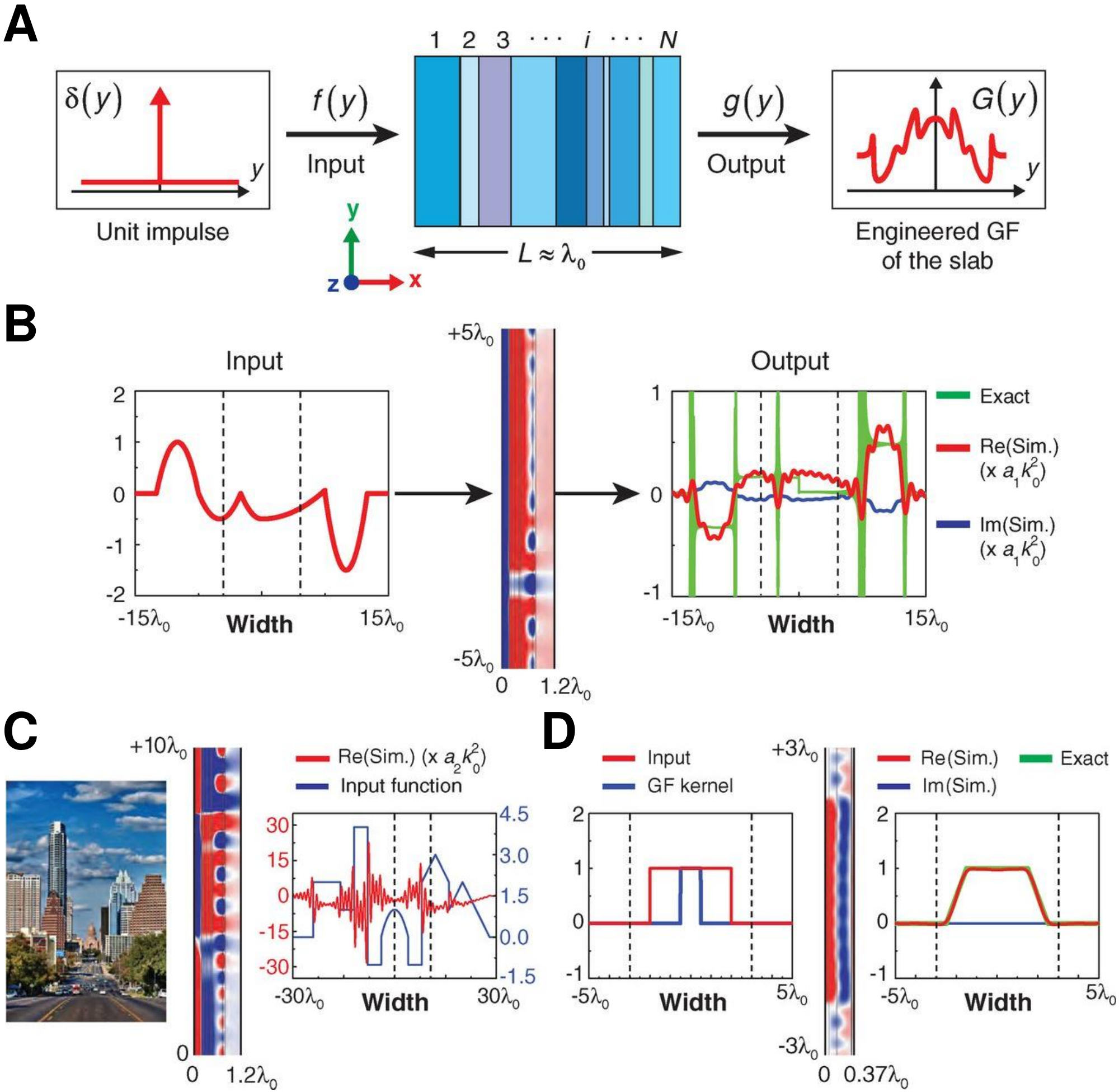}
	\caption{Spatial analog computing using metamaterial slabs based on Green's function approach \cite{silva2014performing}.
	(A) Designed multilayered slab to realize the desired Green's function kernel corresponding to the selected mathematical operation.
	Evolution of electromagnetic wave in the ten-layered nonmagnetic Green's function slab designed to perform second-order differentiation on a quadratic polynomial functions (B) and a city skyline (C). In (C) the simulated output result is directly compared with the input function to highlight the performance of the system in detecting sharp edges.
	(D) Green's function slab with five layers designed to realize convolution with a rectangular kernel.}
	\label{figS8}
\end{figure}

In the GF approach, the desired optical transfer function is directly implemented into the wave vector domain (Fourier or k-space) by using the nonlocal (angular dependent) response of a suitably designed metamaterial or metasurface as shown in Figure~\ref{figS0}B \cite{zhu2020optical, zhou2020analog, cordaro2019high, kwon2018nonlocal, fang2017grating}. Due to the direct implementation of the desired transfer function in this method, performing Fourier/Inverse Fourier transforms are no longer needed, which in turn leads to the reduction in the overall structure size \cite{silva2014performing}. On the other hand, it can be very difficult to realize a desired, but arbitrary, complex transfer function in GF approach. Here, we present a general 2D description of the GF approach which can be simplify into 1D description as described later in this section.  
First, assume that a field component of a monochromatic beam propagating in the $+z$ direction is modulated by an arbitrary 2D signal profile $f(x,y)$. This field profile can be decomposed into TE and TM linearly polarized plane waves when represented in the wave vector space. Through Fourier transform, one can obtain the complex amplitude of each of these plane wave components, i.e. $F_{TE }(k_x, k_y)$ or $F_{TE }(k_x, k_y)$. It is noteworthy that $F_{TE }(k_x, k_y)$ and $F_{TE }(k_x, k_y)$ can be obtained only by knowing the transverse components of the electric field. One can calculate other field components using Maxwell’s equations as described in Ref.~\cite{bykov2014optical}.
The next step is to judiciously design an optical structure that properly manipulates the amplitude and phase of each of the decomposed TE and TM plane wave of the incident field profile. The application of such a desired amplitude and phase on these plane waves can be represented as an Optical Transfer Function (OFT) denoted by \cite{roberts2018optical}:

\begin{equation}
\label{GF_tensor}
\bar{\textbf{H}}(k_x, k_y)= \begin{pmatrix}
H_{TE-TE}(k_x,k_y) & H_{TE-TM}(k_x,k_y) \\
H_{TM-TE}(k_x,k_y) & H_{TM-TM}(k_x,k_y)
\end{pmatrix}.
\end{equation}

The OFT tensor $\bar{\textbf{H}}(k_x, k_y)$ can transform the incident field profile to a reflected or transmitted field profile.  The manipulation of the geometry of the structure representing the $\bar{\textbf{H}}(k_x, k_y)$ allows for designing the on-diagonal elements (coupling of parallel polarizations) and off-diagonal elements (coupling of orthogonal polarizations) in Equation~\ref{GF_tensor}. It should be noted that in this formulation, only the zeroth order reflection and refraction are considered to occur. The elements of the OTF of a certain structure can be calculated using numerical simulation methods such as finite element method (FEM) of finite difference time domain (FDTD) \cite{roberts2018optical}. By simply multiplying the incident field profiles with the OFT of a structure, one can obtain the transmitted or reflected field profile of the corresponding output plane waves as follow \cite{roberts2018optical}:

\begin{equation}
\label{GF_out}
\begin{pmatrix}
G_{TE}(k_x,k_y) \\
G_{TM}(k_x,k_y) 
\end{pmatrix}=\bar{\textbf{H}}(k_x, k_y) \begin{pmatrix}
F_{TE}(k_x,k_y) \\
F_{TM}(k_x,k_y) 
\end{pmatrix}.
\end{equation}

To design the tailored structures for realization of the desired OTFs, different optical phenomena have been used. Here, we classify them into two main categories as resonant and non-resonant based optical phenomena.


As a leading work, Silva $et~al.$ investigated an optimized multilayered metamaterial slabs, which are homogeneous in the transverse directions while inhomogeneous in the longitudinal direction, to realize an appropriate Green's function associated with the second-order spatial derivation \cite{silva2014performing}. A fast synthesis method to calculate the corresponding optical constant and thickness of parallel subwavelength layers was developed. The 10-layer nonmagnetic metamaterial in Figure~\ref{figS8}A is designed to manipulate the transmission coefficient to match the second-order derivative kernel for any incidence angle. Figure~\ref{figS8}B and \ref{figS8}C demonstrates the evolution of magnetic field distribution and the simulation results at the output of the multilayered slab upon excitation with a nonregular polynomial function and a city skyline border, respectively. Figure~\ref{figS8}D shows the simulation results for a five-layer Green's function slab characterized by a rectangular spatial kernel performing convolution operation of a rectangular function.

\subsection{Resonant-based GF approach}

Different types of optical resonances such as Fano resonance \cite{limonov2017fano, luk2010fano, xu2016fano}, Surface Plasmon Resonance (SPR) and Guided Mode Resonance (GMR) have been utilized for implementing OTFs to perform mathematical operations in wave vector domain.

In 2017, Zhu $et~al.$ experimentally demonstrated surface-plasmon-based spatial differentiator using a simple metal-dielectric plasmonics structure in the reflective Kretchmann prism configuration as shown in Figure~\ref{figO1}A \cite{zhu2017plasmonic}. They showed that when the parallel component of the TM-polarized incident wave to the interface matches with the wave vector of the surface plasmon polariton (SPP) at the metal-air interface, a strong SPP is excited and propagates along the metal surface. The interaction between the radiation of the SPP leakage with the direct reflection from the glass-air determine the amplitude of the reflected beam, and in turn, the OTFs of the structure. Considering this mechanism and using coupled mode theory, the authors in Ref.~\cite{zhu2017plasmonic} showed that the OTF of this structure around $k_x=0$ and under critical coupling condition (achieved by simply controlling the thickness of the metal film) can be approximated by:

\begin{figure}[htbp]
	\centering
	\includegraphics[trim=0cm 0cm 0cm 0cm,width=0.5\textwidth,clip]{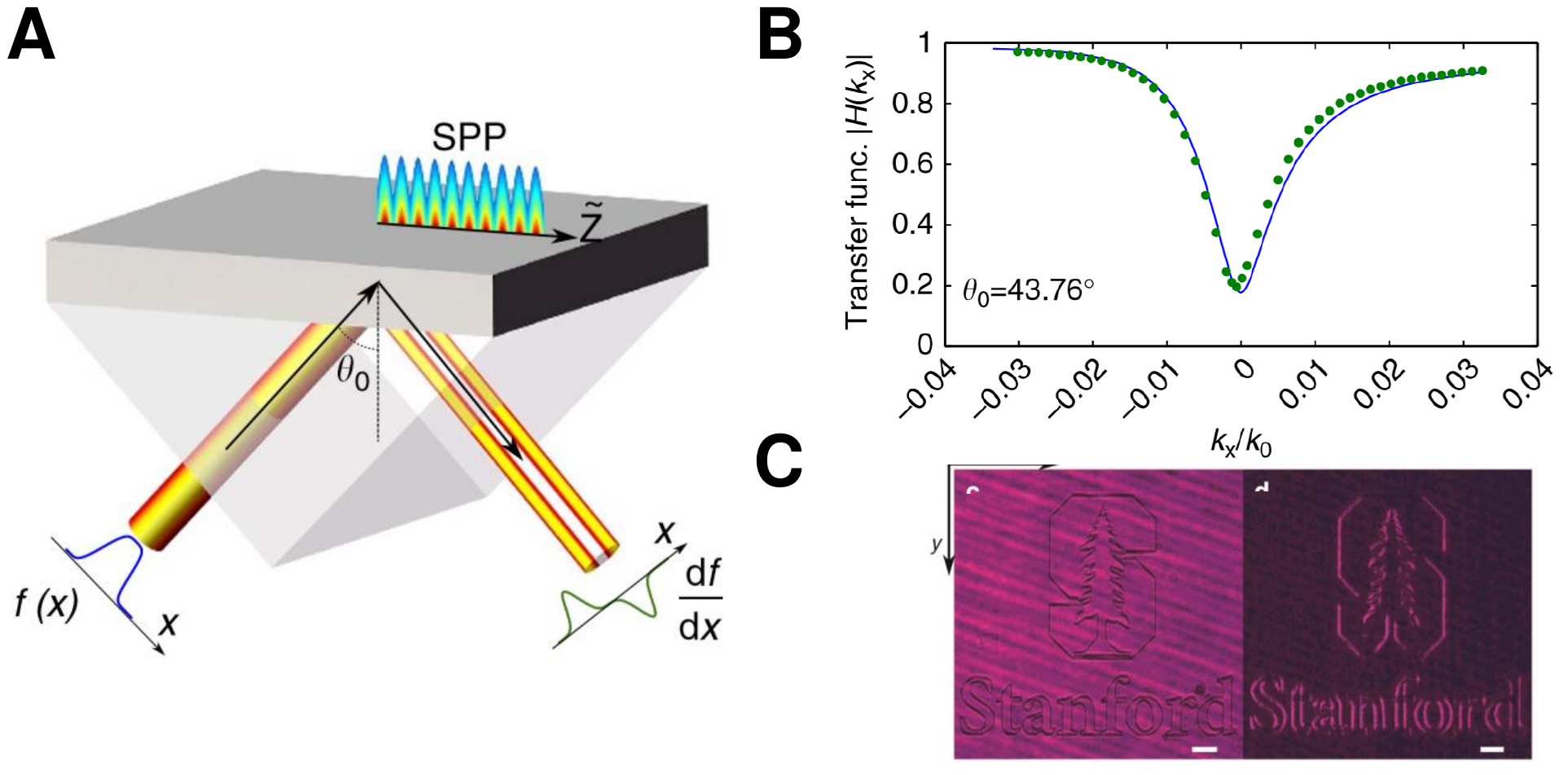}
	\caption{(A) The structure of a surface plasmon-based first-order spatial differentiator consisting of a Ag film on top of the glass in the Kretschmann conﬁguration \cite{zhu2017plasmonic}. (B) The first-order spatial differentiator OTF obtained by the experimental measurement (dotted lines) and the numerical ﬁtting (solid lines). (C) Edge detection of the Stanford tree logo and the letters as the incident image (left panel) using the plasmonic structure shown in (A). The reflected intensity image is shown in the right panel.}
	\label{figO1}
\end{figure}

\begin{equation}
\label{R_out}
\bar{H}_{TM}(k_x) \approx (e^{i \phi}i k_x)/B
\end{equation}
in which $\phi$ corresponds to the phase change during the direct reﬂection at the glass–metal interface and $B$ is expressed in terms of the radiative leakage rate of the SPP and the intrinsic material loss rate, showing that the structure in Figure~\ref{figO1}A can perform first-order derivative for off-normal incidence (i.e., $\theta_0 \ne 0$) as shown in Figure~\ref{figO1}B. Given that a ﬁrst-order differentiator is able to map the abrupt changes in the spectrum (either amplitude or phase) of a input signal to some sharp peaks in an output signal, the authors in Ref.~\cite{zhu2017plasmonic} experimentally demonstrated 1D edge detection of the input Stanford logo image as shown in Figure~\ref{figO1}C.

\begin{figure}[h!]
	\centering
	\includegraphics[trim=0cm 0cm 0cm 0cm,width=0.5\textwidth,clip]{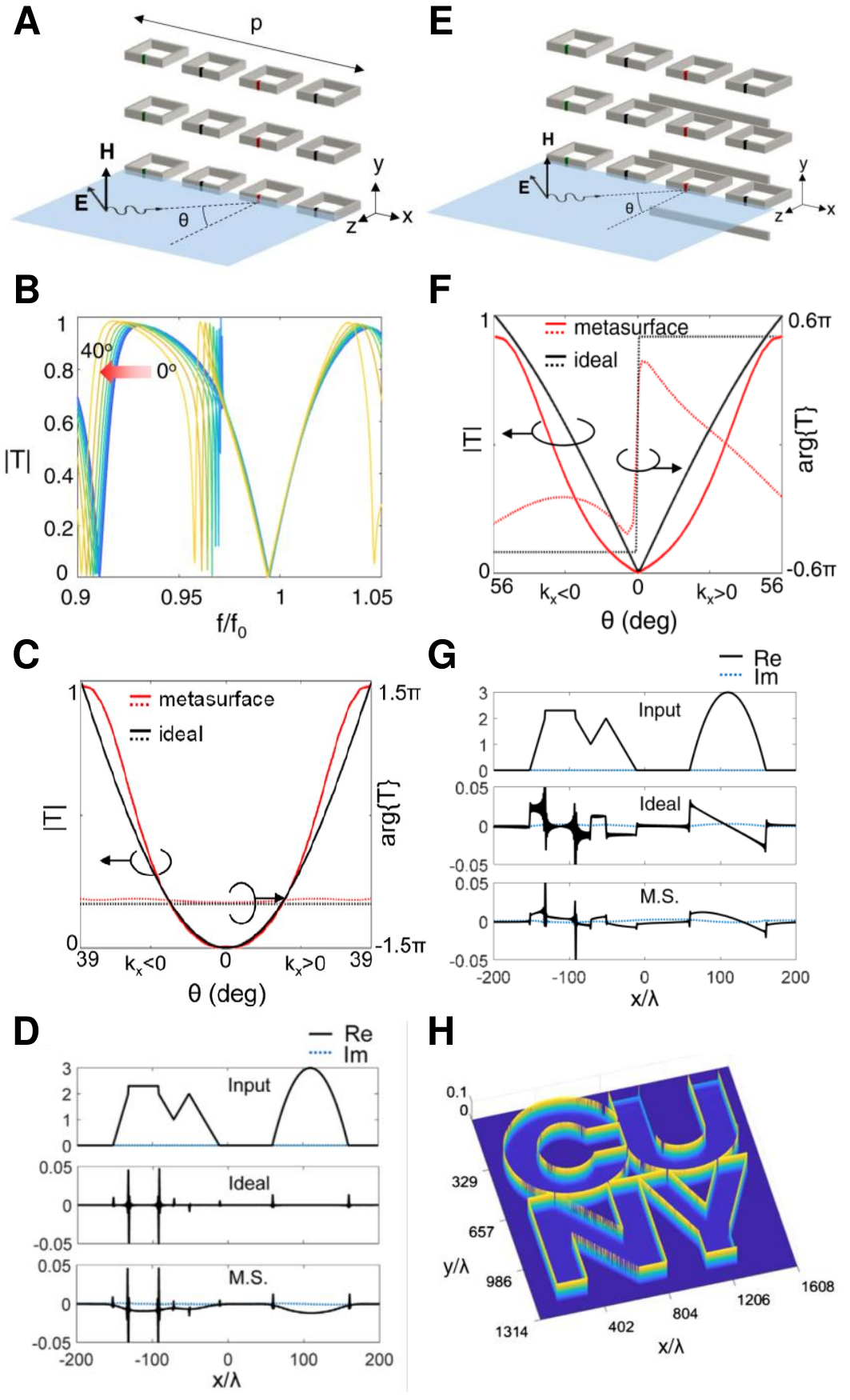}
	\caption{Nonlocal Metasurface. (A) The structure of a metasurface consisting of an array of split-ring resonators (SSRs) under TM illumination \cite{kwon2018nonlocal}. (B) The spectrum of the transmission response of the metasurface when SSRs are periodically modulated. (C) The OTF if the metasurface fitted to a second-order differentiator. (D) Output of an ideal second-order differentiator and of the metasurface from the input signal shown in the top panel. (E) Broken-symmetry metasurface enabling an asymmetric response with respect to positive and negative $k_{x}$ to realize first-order differentiation. (F) The transmission profile as a function of the incidence angle for a symmetry-broken metasurface enabling first-derivative operation. (G) Output of an ideal first-order differentiator working as an edge detector. (H) Detected edges of the CUNY logo for illumination with unpolarized light from the normal direction \cite{kwon2018nonlocal}.}
	\label{figO2}
\end{figure}

\begin{figure}[htbp]
	\centering
	\includegraphics[trim=0cm 0cm 0cm 0cm,width=0.5\textwidth,clip]{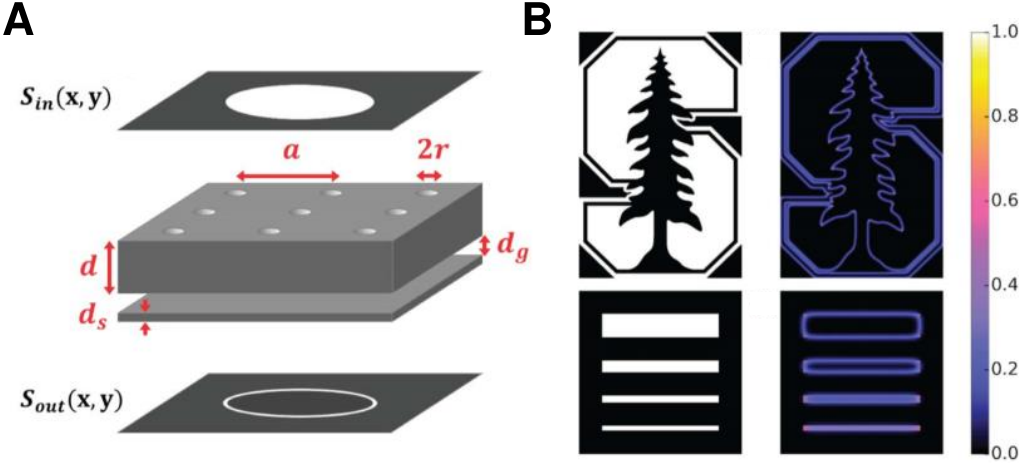}
	\caption{(A) The schematic of the photonic crystal structure for Laplacian differentiation consisting of a photonic crystal slab separated from a uniform dielectric slab by an air gap \cite{guo2018photonic}. (B) Incident Stanford emblem (top-left panel) and  calculated transmitted image with unpolarized light (top-right panel). The spatial resolution of the differentiator is shown in the bottom panels (incident slot patterns in the left, and calculated transmitted images with unpolarized light in the right).}
	\label{figO3}
\end{figure}

To broaden the operational spatial bandwidth for on-demand high resolution edge detection, Kwon $et~al.$ proposed to use a high-efficiency transmittive metasurface consisting of a periodic array of split-ring resonators (SRRs) as demonstrated in Figure~\ref{figO2}A \cite{kwon2018nonlocal}. They theoretically showed that by introducing a periodic (sinusoidal) permittivity modulation in SRR gaps, the nonlocal response of the metasurface can be engineered to achieve desired OTFs for the second-order differentiator. The interaction of the surface GMRs with the leaky-wave resonances can lead to some Fano resonances in the transmission response of this metasurface as shown in Figure~\ref{figO2}B. Working at the transmission zero frequency of the Fano resonance at normal incidence, the transmission response of the nonlocal metasurface changes from zero to unity for oblique incidence (see Figure~\ref{figO2}C). This functionality can be utilized for performing wide bandwidth second-order differentiation as shown in Figure~\ref{figO2}D. They also modified this structure by adding a misplaced array of metallic wires to break the vertical and horizontal mirror symmetry (see Figure~\ref{figO2}E). Such a suitably engineered configuration enables realization of the first-derivative kernel as shown in Figure~\ref{figO2}F. Figure~\ref{figO2}G represents the input signal and the corresponding simulated response of this mathematical operator in caparison to the numerically calculated result. In a further study, by integration of two identical 1D-operation metasurfaces relatively rotated by $90^{\circ}$, which implies $90^{\circ}$ rotational symmetry, such a configuration enabled realization of an identical second-order derivation for a TM-polarized waves along the x and y axes. This platform can be leveraged for more practical 2D edge detection scenarios where unpolarized light for the illumination of the image is utilized as depicted in Figure~\ref{figO2}H where the edges of the CUNY logo are well resolved.

The authors in Ref.~\cite{saba2018two} employed the GMR of a periodic array of dielectric resonators buried in a dielectric slab to perform 2D edge detection. The interaction between the GMRs with the leaky waves of the structure results in some Fano resonances in the transmission response \cite{peng1996resonant}. Working at the zero transmission corresponding to the normally incident light, the authors numerically demonstrated a second-order differentiator owing to the even symmetry of the structure. In another work, the authors in Ref.~\cite{bykov2018first} experimentally demonstrated a first-order differentiator by employing GMRs in the case of an oblique incident Gaussian beam. GMRs have also used for implementing optical integrators. The authors in Ref. \cite{zangeneh2017spatial} utilized a prism coupling configuration to implement the OTF of a 1D optical integrator in the transmission response of a multi-layer dielectric slab structure. It is shown that the dielectric slabs can be replaced by graphene sheets to enable miniaturized optical integrator.

In another work, Gue $et~al.$ proposed to use a photonic crystal slab (see Figure~\ref{figO3}A) in the tranmission mode to realize the Laplacian operation on a 2D input, i.e. $\nabla^2 = \partial^2_x + \partial^2_y$. The $\nabla^2$ operation in real space is equivalent to the following OTF \cite{guo2018photonic}:

\begin{equation}
\label{guo}
\bar{\textbf{H}}(k_x, k_y)= \begin{pmatrix}
T_{TE-TE}(k_x,k_y) & 0 \\
0 & T_{TM-TM}(k_x,k_y)
\end{pmatrix}.
\end{equation}

\begin{figure}[htbp]
	\centering
	\includegraphics[trim=0cm 0cm 0cm 0cm,width=0.5\textwidth,clip]{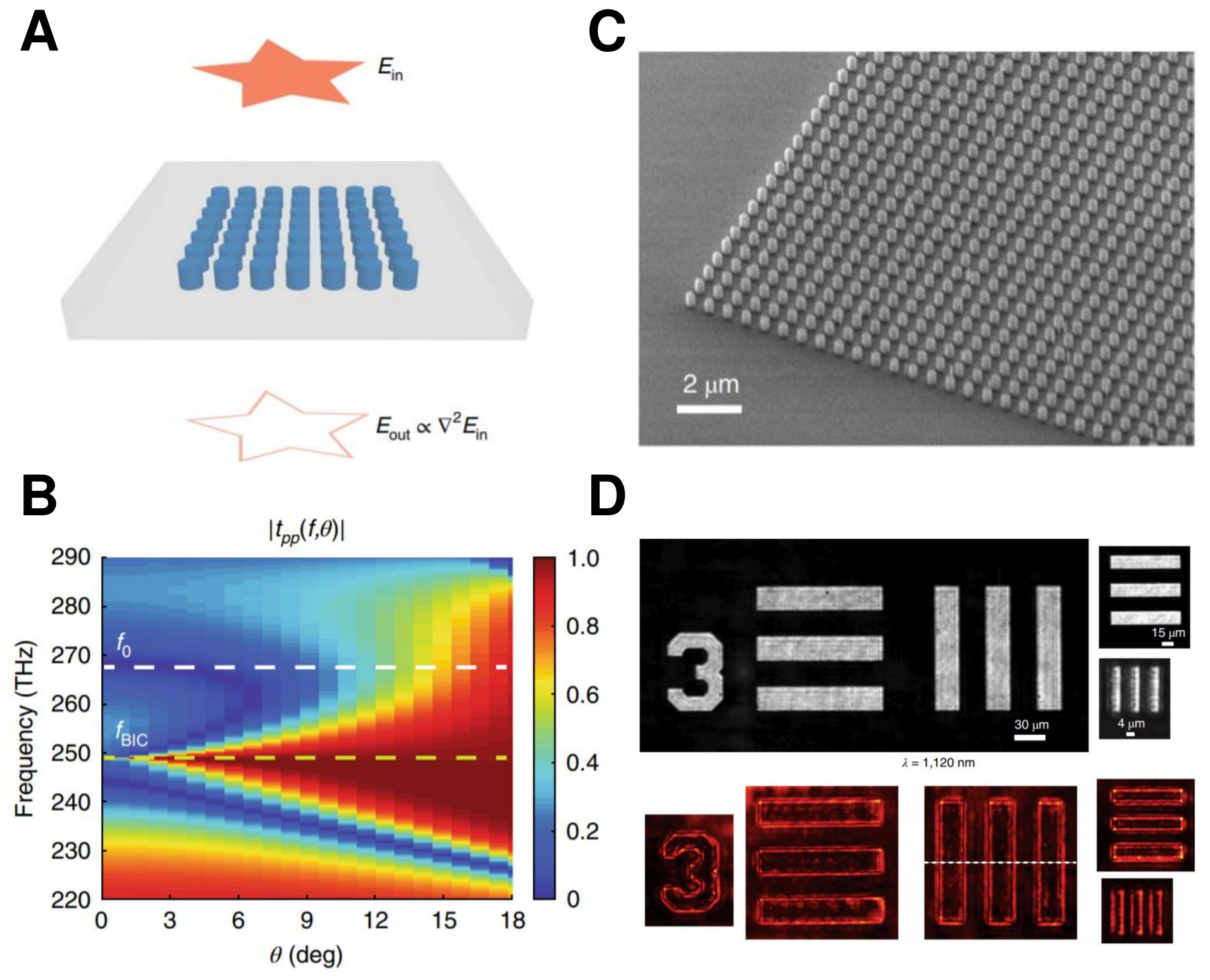}
	\caption{2D image differentiation using nanophotonic materials \cite{zhou2020flat}.
	(A) Schematic of a photonic crystal slab acting as a Laplacian operator that transforms an image, $E_\textrm{in}$, into its second-order derivative, $E_\textrm{out} \propto \nabla^{2}E_\textrm{in}$.
	(B) Simulated colour-coded transmission coefficient amplitude as a function of frequency and incident angle along the $\Gamma-X$ direction ($\phi = 0^{\circ}$) for p polarization.
	(C) SEM image of the fabricated Si photonic crystal.
	(D) Imaging results for the target without (top row) and with (bottom row) the differentiator.
	}
	\label{figS11}
\end{figure}

To realize this ideal 2D Laplacian operation, the authors in Ref.~\cite{guo2018photonic} utilized the guided modes of the photonic crystal slab near the $\Gamma$ point in the Brillouin zone, and the authors in Ref.~\cite{guo2018isotropic} used isotropic image filters to have identical response to both polarization, and make the off-diagonal elements equal to zero as required by Equation~\ref{guo}. It is worth mentioning that working in the transmission mode is more compatible for image processing applications which is an advantage of the work in Ref.~\cite{guo2018photonic} over other approaches for Laplace operator implementation \cite{bykov2014optical}. The numerical demonstration of the proposed structure shown in Figure~\ref{figO3}A is demonstrated in Figure~\ref{figO3}B where the transmitted image of an unpolarized incident beam of the Stanford emblem as well as some slot patterns are calculated. Later, Bezus $et~al.$ employed the resonance of a dielectric ridge on a slab waveguide to implement both optical differentiation and integration in the spatial domain \cite{bezus2018spatial}.

\begin{figure}[htbp]
	\centering
	\includegraphics[trim=0cm 0cm 0cm 0cm,width=0.5\textwidth,clip]{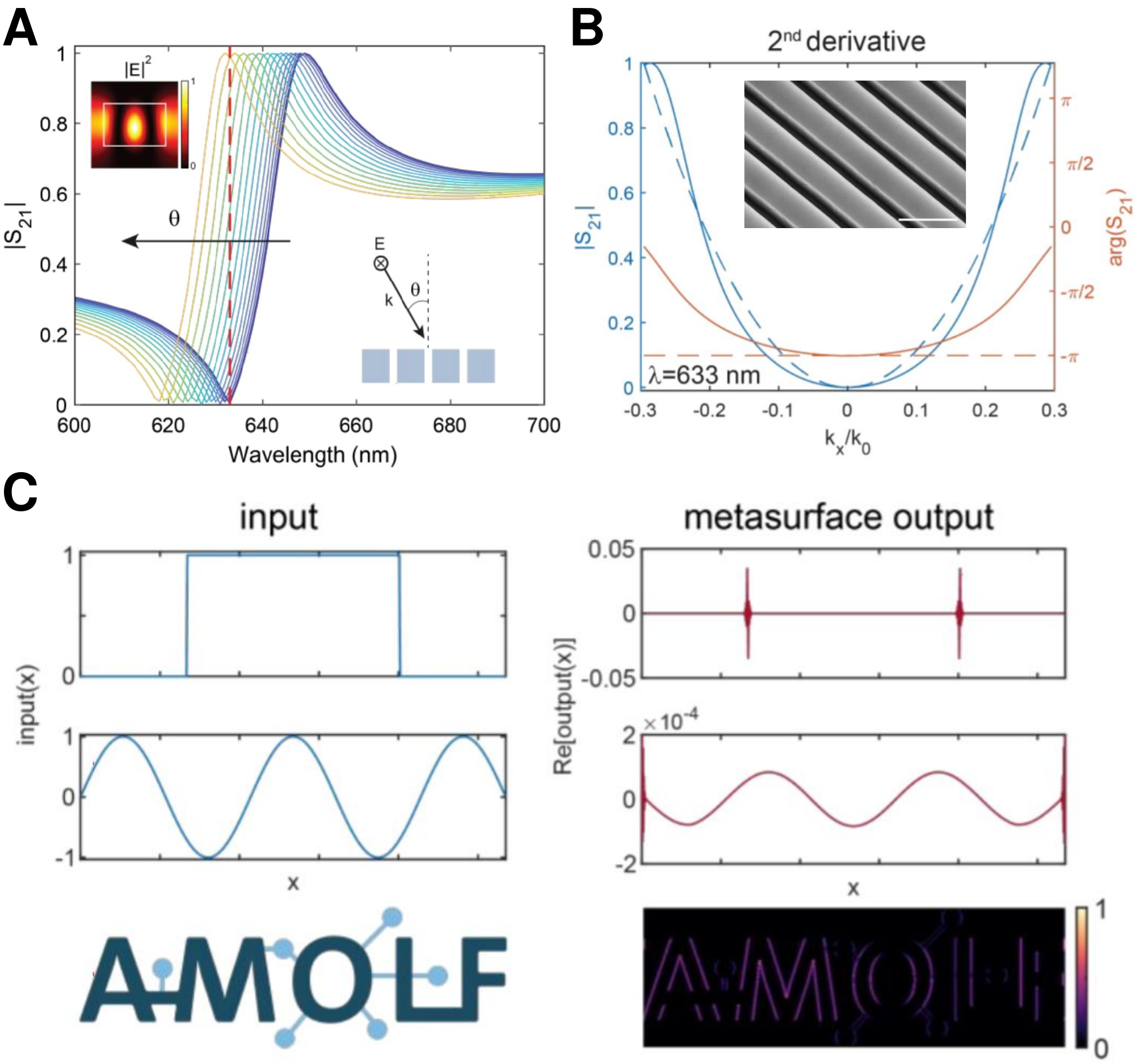}
	\caption{Optical computing based on Fano-resonant metasurface \cite{cordaro2019high}.
	(A) Simulated transmission spectra of a metasurface consisting of an array of dielectric nanobeams as the incident angle is changed from 0 (blue line) to 0.3 rad (yellow line) in 15 steps. The red dashed line indicates the wavelength of operation (633 nm).
	(B) Transmission amplitude (solid blue line) and phase (solid orange line) for the metasurface optimized for second-derivative operation (sketched in the inset) at 633 nm. The simulated transfer function is compared to the ideal case (dashed lines). The transmission reference plane is set such that the transmission phase at normal incidence equals $-\pi$.
	(C) Rectangular and sinusoidal input functions and 2D image that are used to numerically test the metasurface operation. The signal is discretized into 1000 pixels with individual pixel size set such that the Nyquist range matches the operational range in k-space of the metasurface (left column). Metasurface output for the input in part b. For the 2D image, differentiation is performed line by line along the x-axis (right column). The size of each input pixel is set so that the Nyquist range matches the operational bandwidth of the metasurface in the k-space.}
	\label{figS10}
\end{figure}

Valentine's group, more recently, demonstrated the applicability of flat optics for direct image differentiation using a compact all-dielectric meta-structure \cite{zhou2020flat}. In their work, first a metasurface-based differentiator was used in combination with the conventional optical components and a camera sensor for high-speed edge detection (see Figure~\ref{figS11}A). Such a metasurface afford improved transmission amplitude profile as shown in Figure~\ref{figS11}B. In a second approach, a Si-based photonic crystal in combination of a metalens were employed to realize the second-order derivation to directly discriminate edges in an image (see Figure~\ref{figS11}C). Such a compact and monolotic image processing system paves the way for real-time computer vision tasks, as depicted in Figure~\ref{figS11}D.


In a follow up work to Ref.~\cite{kwon2018nonlocal}, Cordaro $et~al.$ designed and implemented an all-dielectric metasurface formed by a 1D array of Si nanobeams to locally tailor the optical properties of the incident S-polarized light \cite{cordaro2019high}. Due to engineered spatial dispersion, the interference of the broad Fabry-Perot resonance of structure with the quasi-guided mode along the surface results in an asymmetric Fano resonance in the transmission mode Figure~\ref{figS10}A. The sharp response in frequency is associated with strong nonlocality governing the angular sensitivity of the transmission response. The strong amplitude modulation correlated to the incoming k-vector is the basis for implementing a spatial filter associated with the mathematical operator of choice. The simulated transmission amplitude and phase as a function of the normalized in-plane wavevector for the second-order differential kernel at the operational wavelength of 633 nm is shown in see Figure~\ref{figS10}B. As shown in see Figure~\ref{figS10}C the edges of the rectangular input profile and the flip of the sinusoidal input function are appeared as the output of the simulated metasurface.

\begin{figure}[t]
	\centering
	\includegraphics[trim=0cm 0cm 0cm 0cm,width=0.5\textwidth,clip]{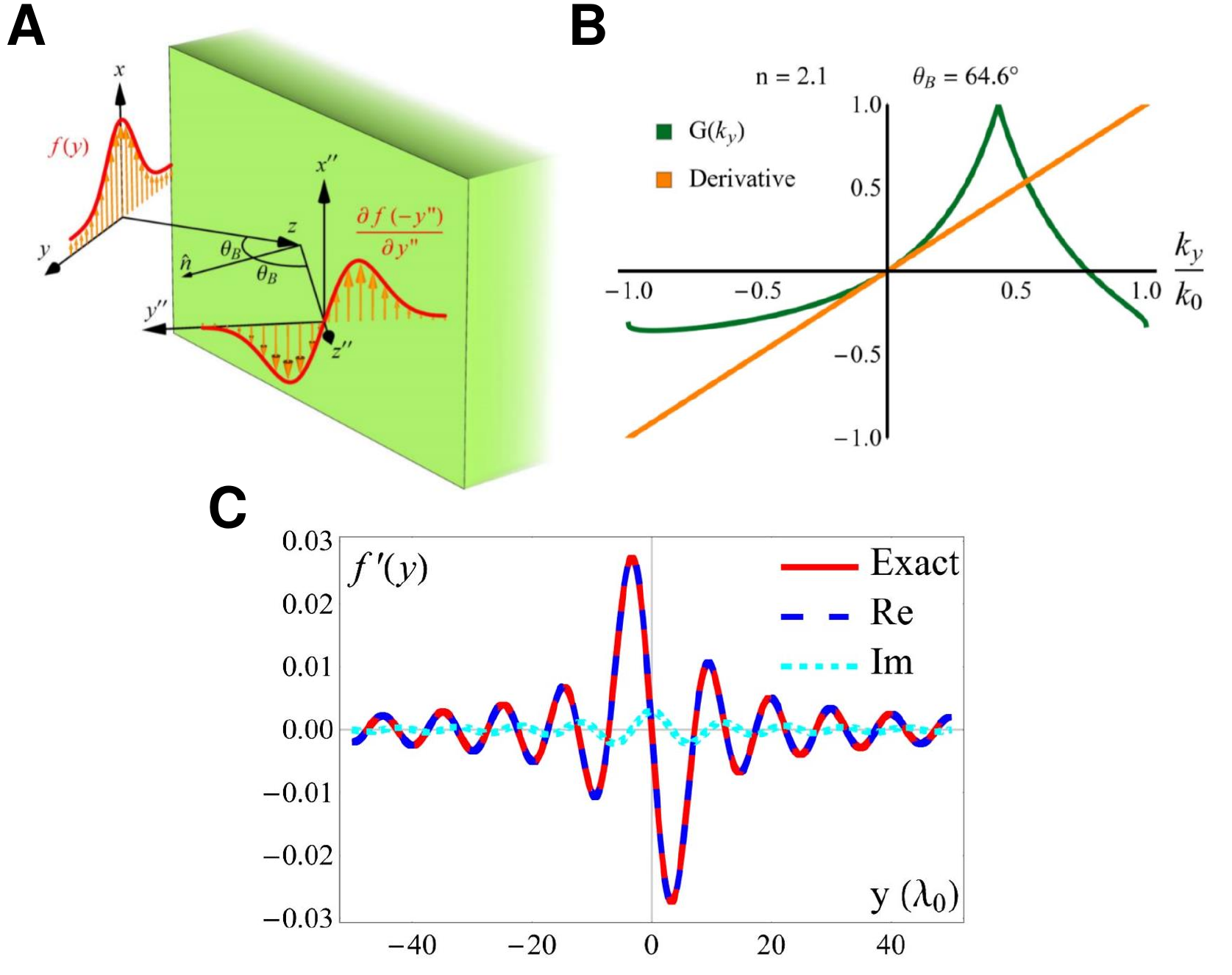}
	\caption{Example of non-resonant-based GF approach. (A) The realization of the first-order differentiation when an incident beam impinging and reflected at Brewster angle \cite{youssefi2016analog}. (B) The exact and approximated GF distribution corresponding to the first-order differentiation. (C) Comparison of the exact solution and the simulated first-order derivation for an input Sinc function with $W = 0.09 \, k_{0}$.}
	\label{figO4}
\end{figure}

Several other proposals and demonstrations using spin Hall effect of light \cite{zhu2019generalized}, prism coupling configuration \cite{zhang2019time},
reflective hybrid plasmonic-dielectric metasurfaces \cite{zhou2020analog}, periodic plasmonics metasurfaces covered by graphene \cite{lou2020optical}, multi input-multi output computational metasurface \cite{babaee2020parallel}, ultra-thin bianisotropic metasurfaces \cite{abdolali2019parallel}, and polarization-insensitive structured surface with tailored nonlocality \cite{kwon2020dual} to perform mathematical operators based on Green's function approach have been reported in the literature.

\subsection{Non-resonant-based GF approach}

The authors in Ref.~\cite{youssefi2016analog} employed the Brewster effect, which is associated with the zero reflection of a TM-polarized incident beam from an interface between two dielectric medium, to realize first-order differentiation. The schematic of their proposed structure is shown in Figure~\ref{figO4}(A) in which the symmetry of the system is broken by applying oblique incident beam. Figure~\ref{figO4}(B) shows the exact GF and its approximation around $k_y=0$ which can be used for implementation of first-order derivative. To study the performance of the proposed configuration, the proposed structure is illuminated with a Sinc function beam profile with the bandwidth of $W = 0.09 \, k_{0}$ at the Brewster angle. The calculated first-order derivative of the input field is compared with the exact solution as shown in Figure~\ref{figO4}(C). Later, the same group demonstrated that using a simple half-wavelength dielectric slab in the reflection mode, ﬁrst-order differentiation can be realized \cite{zangeneh2018analog}. Using a transmission-line approach, they showed that employing impedance matching condition, one can perform ﬁrst-order differentiation on oblique incident input signals.

\begin{figure}[t]
	\centering
	\includegraphics[trim=0cm 0cm 0cm 0cm,width=0.5\textwidth,clip]{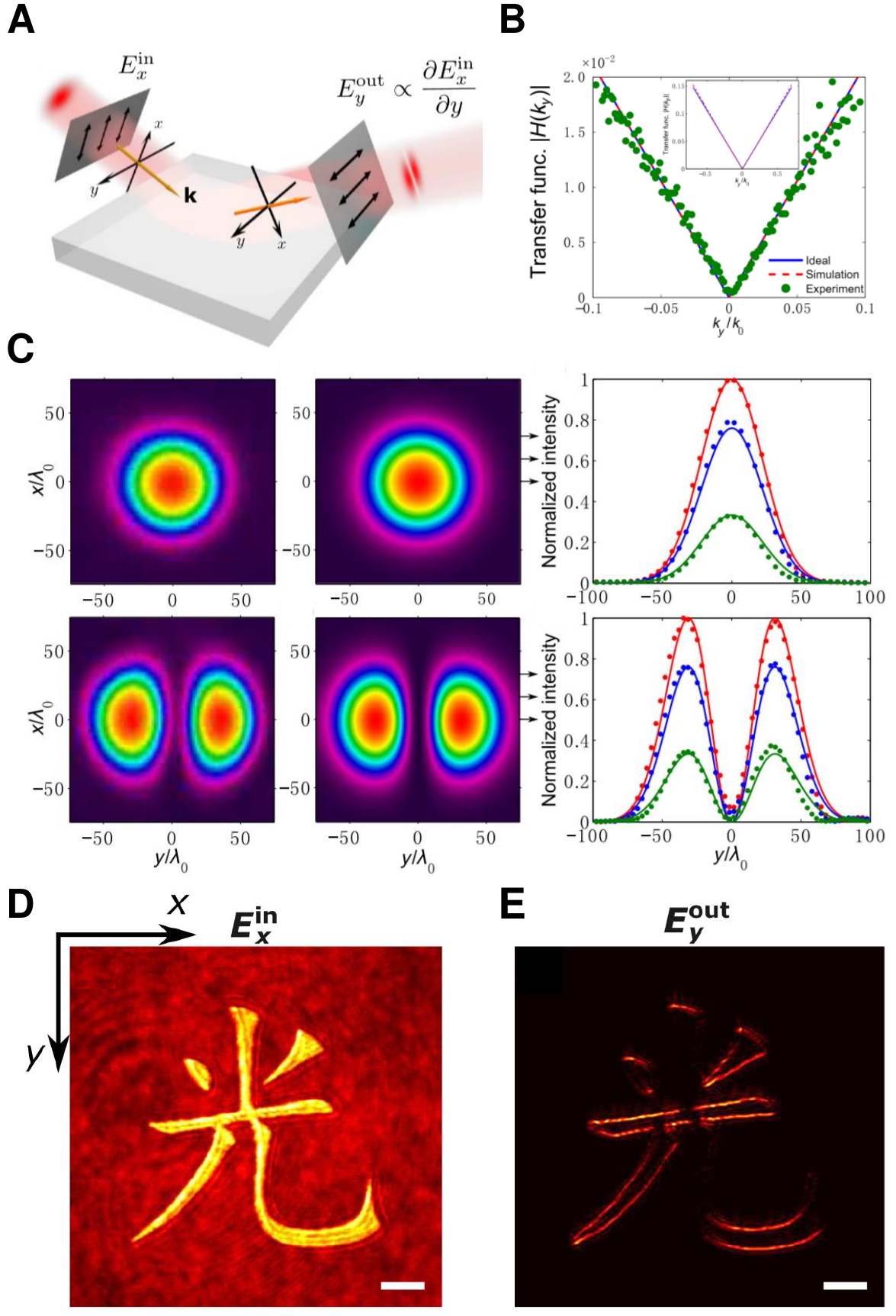}
	\caption{(A) Schematic of spatial diﬀerentiation from the spin Hall effect (SHE) of light on an optical planar interface between two isotropic materials, e.g., an air-glass interface \cite{zhu2019generalized}. (B) Measurement of the spatial spectral transfer function on an air-glass interface. (C) Spatial diﬀerentiation demonstration for a Gaussian illumination with an incident angle $\theta_0=45^\circ$. (D) Incident image comprises a Chinese character for encoded with amplitude modulation on $E^{\textrm{in}}_{x}$. (E) Measured cross-polarized reflected light (i.e., $E^{\textrm{in}}_{y}$) form the structure illuminated with the incident filed in (D).}
	\label{figO5}
\end{figure}

Recently, Zhu $et~al.$ experimentally demonstrated spatial differentiation of the incident beam when reflected or refracted at a single optical planar interface under paraxial approximation \cite{zhu2019generalized}. They employed spin Hall effect (SHE) of light, i.e. the polarization-dependent transverse shift of an optical beam totally reflected from a planar interface \cite{hermosa2011spin}, to compute the spatial differentiation of the incident beams. As Figure~\ref{figO5}C shows, when the obliquely incident paraxial beam has an electric field distribution of $f(x,y)$, the output field distribution associate with $df(x,y)/dy$. To validate their finding, the authors in Ref.~\cite{zhu2019generalized} performed experimental measurements of the OTF of a glass-air interface as shown in Figure~\ref{figO5}C. Moreover, they applied their method to a Gaussian incident beam profile and performed the first-order spatial differentiation as the ﬁrst-order Hermite-Gaussian proﬁle of the reflected beam in Figure~\ref{figO5}C shows. Finally, they applied their spin-optical method to perform an edge detection operation. Figure~\ref{figO5}D demonstrates the incident image field of a Chinese character encoded in the x-component of the field based on the amplitude modulation. The measured y-component of the reflected field exhibits the resolved outlines of the character (see Figure~\ref{figO5}E) which are more visible in the x direction. The edges parallel to the x direction are more visible since the differentiation is performed along the y direction.


\begin{figure}[h!]
	\centering
	\includegraphics[trim=0cm 0cm 0cm 0cm,width=0.5\textwidth,clip]{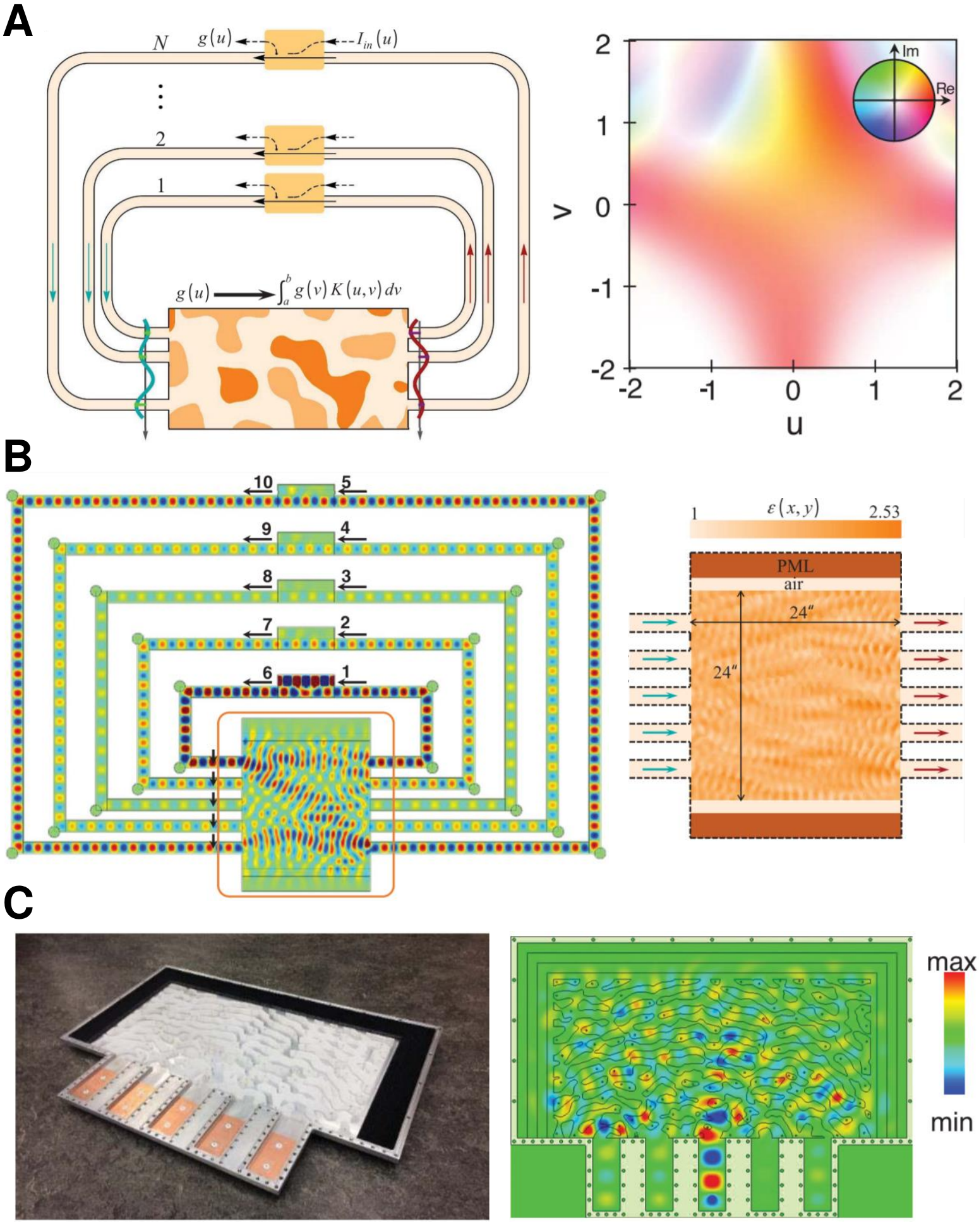}
	\caption{Solving integral equation with waves in inversve-designed meta-structures \cite{estakhri2019inverse}.
	(A) Left panel: the schematic of a closed-loop network consisting of a properly designed kernel operator (i.e., the metamaterial block) that performs the desired integral operation on the input wave, N feedback waveguides for realization of the recursive paths, and directional couplers to feed/probe the input/output wave. Right panel: complex density plot of the first complex-valued kernel $K_{1}(u, v)$ in the $(u, v)$ plane in Equation~2 in \cite{estakhri2019inverse}. (B) Left panel: time snapshot of the simulated out-of-plane component of the electric field distribution in the closed-loop network when excited at port 1, with the other four ports unexcited. Right panel: inhomogeneous metamaterial kernel designed for performing integral operation enclosed with absorbing layers on top and bottom, and perfect electric conducting walls on all other sides.
	(C) Left panel: an image of the fabricated meta-structures in which white, gray and black areas show polystyrene, air and absorbing materials, respectively. The outer electric conducting walls are made of aluminum. Right panel: time snapshot of the simulated out-of-plane component of the electric field distribution in the closed-loop network when excited at port 3.}
	\label{figS9}
\end{figure}

More recently, Estakhri $et~al.$ in a disruptive approach demonstrated a computational meta-structure platform that solves linear integral equations by tailoring the complex-valued electromagnetic wave propagating through the specially designed recursive paths \cite{estakhri2019inverse}. The conceptual representation of the meta-structure is shown in Figure~\ref{figS9}A where the solution to the Fredholm integral equation of the second kind is obtained. While the given integral operator with kernel $K(u, v)$ is implemented in the metamaterial block, the input signal $I_\textrm{in}(u)$ is introduced to the equation through a set of coupling elements along the feedback waveguides. The overall relation describing the behavior of the equation solver follows $g(u) = I_{in}(u) + \int_{a}^{b} K(u,v)g(v)\textrm{v}$, in which $g(u)$ is the unknown solution of the integral equation. To realize the equivalent $N \times N$ matrix equation of the abovementioned governing equation, $N$ feedback waveguides are exploited to sample the input of the metamaterial block to its output. To showcase the performance of this approach in a physical system, the authors leveraged a judiciously patterned meta-structure included in a feedback loop, which calculated the inverse of a known $N \times N$ matrix. Figure~\ref{figS9}B shows the numerical simulation results for the distribution of the electric field in the assembled system consisting of the prescribed meta-structure in combination with the five feedback waveguides and directional couplers. The comparison assessment between the theoretical and simulation results corroborate the fidelity of the proposed approach. In a proof-of-concept demonstration at the microwave regime, the feedback process is internally performed in a reflective system, as shown in Figure~\ref{figS9}C. Topology optimization method was leveraged to inversely design an optimized meta-structure realizing the kernel associated with the operator of choice. The simulated electric field and comparison between the numerical and experimental results are represented in Figure~\ref{figS9}C.

\section{Summary and Outlook}

Spatial analog computing platforms facilitate wave-based, real-time, high-throughput, and large-scale information processing with low-energy consumption. Fourier transformation and Green's function have been pursued as two powerful concepts for realization of mathematical operations. While the former is relied on 4\textit{f} systems with two Fourier transform lenses and one intermediate spatial frequency filter, the latter leverages the nonlocal response of judiciously designed optical component to implement the desired kernel. Due to their unprecedented capabilities in sculpting the scattered light, meta-structures enable miniaturization of traditional bulky optical systems to the integrable ultra-compact frameworks. The intersection of the aforementioned fundamental theoretical concepts and functionalized metamaterials and metasurfaces promise realization of computational meta-optics. We discussed recent advances in spatial analog optical computing devices enabling mathematical operations including integration, differentiation, and integro-differential equations solution as well as spatial frequency filters. Despite the existing promising platforms and design approaches enabling computational meta-systems, there is still room for improvement and progress in this rather infant field. In what follows, we elaborate some of unexplored aspects, key challenges, and possible opportunities in the general field of computational meta-optics.\\
In general, the architecture of a conventional optical processor is relied on cascaded input, processing, and output planes. To leverage the profound potential of meta-optics for realization of flat optical processors, miniaturization and possible stacking of planes are indispensable. The first (or input) plane, where mostly an electrical to optical conversion of raw data is performed at the video rate using a SLM, is considered to be the bottleneck of most practical computational systems. More recently, several tuning  mechanisms including electrical, thermal, mechanical, and optical have been introduced \cite{taghinejad2019all} enabling reconfigurable nanoscale SLMs and phased-array antennas. Tuning the refractive index of Si based on thermo-optic effects \cite{wang2020fast}, manipulating the the electro-optic characteristics of indium tin oxide \cite{kafaie2018dual,shirmanesh2019electro} relying on free carrier effects, controlling the effective physical state of liquid crystal \cite{li2019phase} exploiting phase transition effects, exploit the ultrafast transport dynamics of hot electrons in a hybrid crystal \cite{taghinejad2018ultrafast}, and phase conversion of phase-change materials between the amorphous and crystalline states \cite{tittl2015switchable, li2016reversible, abdollahramezani2018reconfigurable,wuttig2017phase,ding2019dynamic, abdollahramezani2020tunable, taghinejad2020ito} have shown interesting capabilities. Such approaches ultimately can grant moderate to high-speed addressability at the pixel-level necessary for high-resolution on-demand imaging and computing applications. Furthermore, by combination of an arrangement of high quality factor resonant nanoantenna with quantum dots, strong spontaneous emission occurs leading to light-emitting metasurfaces. The judiciously designed array of such metasurfaces can effectively manipulate the radiation pattern that paves the way toward creation of on-chip flat sources with tailored light fields \cite{ha2018directional,ma2019applications,vaskin2019light}. The second (or processing) plane which generally accommodates lenses, holograms, or nonlinear elements can also take advantage of best-of-breed all-dielectric metasurfaces. Recent advances in achromatic focusing \cite{chen2020flat}, computational imaging \cite{colburn2018metasurface}, and multi-plane holography \cite{jiang2019metasurface} promise the emergence of these platforms as new paradigms in computational systems. Moreover, recent development of multiplexed and multifunctional metasurfaces through integration of different information channels into a single or multilayered metasurfaces with segmented or interleaved meta-atoms \cite{chen2020metasurface,chen2019single} could have deep impact on the future optical computation technology. Finally, integration of resonant metasurfaces with III-V compound semiconductors with strong adjustablity helps shrinking the overall size of the photodetector arrays or even cameras in the last (or output) layer of an optical computing system.\\
Feature selection and object detection are key steps in image processing applications which heavily rely on edge detection of input data. So far, most demonstration have suffered from the narrow operational spatial bandwidth associated with the low resolution of processed images. In a theoretical study, Karimi \textit{at al.} discussed the fundamental gain-resolution limit that applies to optical analog edge detectors \cite{karimi2020fundamental}. Considering the Rayleigh criterion, the derived formula reveals a linear relation between the gain of the spatial filter and the achievable image resolution. The authors demonstrated that a simple dielectric slab waveguide can theoretically outperform the available flat optical differentiator. The calculated physical bound for a generic spatial differentiator has provided a useful metric to compare and assess the performance of existing edge detection devices, which needs to be taken care of for future improvements. In a recent work, the effect of a time modulated incident light on  the response of a plasmonic spatial differentiator based on a prism coupling configuration is investigated \cite{zhang2019time}. The authors in Ref.~\cite{zhang2019time} have shown that the plasmonic spatial differentiator can operate with an estimated speed of $10^{13}$ frame/s. This high time-bandwidth of the plasmonic spatial differentiator in conjunction with its high space-bandwidth provides the ability for realization of high-throughput real time image processing applications.\\
To improve the robustness of computational metasurfaces against defects or discontinuities at the interface, Zhang \textit{et al.} proposed to use precise tuning of the balance between the asymmetric leak rate and the intrinsic loss rate of the unidirectional SPP leaky mode in a nonreciprocal plasmonic platform \cite{zhang2019backscattering}. To do so, they leveraged the nontrivial topological properties of a gyrotropic material whose time-reversal symmetry is broken under the a static magnetic field. The investigated double-layer flat structure enables ideal realization of first-order spatial differentiation for fairly insensitive edge detection in the terahertz regime. \\
So far, implementation of computational meta-devices have been majorly relied on intuition-based approaches where meta-structures utilizes simple geometries governed by well-understood physics. 
When it comes to mathematical operators governed by complex multifunctional kernels, traditional design techniques (such as brute-force) suffer from considerable drawbacks in translating the well-formulated transfer function to the real physical device platform. 
This is mainly due to hyper-dimensional nature of the optimization problem associated with sophisticated input to output mapping where parametric sweeps exhibit inefficiency to be employed as an ideal optimizer.
Recently, inverse design approaches based on local and global optimization techniques have attracted significant attentions in enabling nontrivial high-performance meta-optic configurations targeted to a wide range of applications. 
Among several inverse design approaches, global step-by-step searching algorithms (such as genetic or particle swarm), adjoint-based topology optimization implementations, and neural network-assisted optimization approaches have proven to be compelling candidates to push forward high-performance nanophotonic devices \cite{jensen2011topology,molesky2018inverse,campbell2019review, liu2018training, lin2018all, liu2018generative, zhan2019controlling, jiang2019free, kiarashinejad2019deep, khoram2019nanophotonic, kiarashinejad2019knowledge, jiang2020deep, kiarashinejad2019deepdim, kudyshev2020machine, an2020freeform, sell2017periodic, khoram2019nanophotonic}.
Coupled to recent advances in nanofabrication technologies, physical modeling, and computational power, such single and multi-objective optimization approaches can benefit next generation computational meta-processors. \\
Thus far, fabrication of functional nanophotonic devices inverse-designed by a topology optimization technique have been relied on high-resolution electron-beam lithography. Despite successful experimental demonstrations, none of these structures have been broadly adopted to the industry mainly due to small features in their structures, which are challenging to be resolved with available industry-standard photolithography technologies \cite{molesky2018inverse}. To address this issue, fabrication constraints have to be incorporated in inverse design algorithms. One way to do so is simply subdividing the design region into rectangular pixels which are larger than the minimum allowable feature size. However, this pixel-based approach is not optimal for most practical optical devices with smooth curves in their structures. Although one approach to consider these smooth curves in the design process is to use a convolution filter followed by thresholding \cite{elesin2012design}, this can result in artifacts smaller than achievable feature sizes. As a solution to this problem, Piggott. \textit{et al.} proposed to impose curvature constraints on the device boundary in the algorithm \cite{piggott2018automated}. In this boundary parametrized optimization, a minimum radius of curvature is enforced to avoid features smaller than an acceptable threshold, and simultaneously, any gaps or bridges narrower than a chosen threshold are periodically eliminated by cutting them in half. Although these techniques have shown improvements in terms of fabrication tolerances when tested through experiments for electron-beam lithography, and through simulations for photolithography, their robustness against systematic/random errors in photolithography, such as defocusing and dosage errors is still questionable. Therefore, it is essential to incorporate the robustness against these fabrication process variations into the optimization algorithm in order to successfully deal with the industry-based fabrication challenges.

Leveraging highly confined SPPs can facilitate the ultra-compact computational devices based on 2D planar plasmonic configurations. Kou $et~al.$ demonstrated the complex operation of Fourier Transform carried out at a velocity close to the speed of light based on highly confined plasmon waves \cite{kou2016chip}. Thanks to the reduced dimensionality, their SPP-based device enables ultracompact computational devices with high accessible spatial resolution way beyond the diffraction limit of the incident light.

Light-matter interaction in 2D materials and transition metal dichalcogenides in the form of plasmon- or exciton-polaritons have attracted intense interests recently \cite{wang2012electronics,novoselov20162d,taghinejad2020photocarrier}. Thanks to their atomic scale size, a great deal of opportunities can be seen for analog computing applications by harnessing these types of dynamics. In particular, easy-to-fabricate, wafer-scale exfoliated graphene can be served as a real deep subwavelength metasurface platform with profound addressability of gate-tunable meta-atoms. Although the expanded family of these materials provides a rich set of metasurface frameworks working at operational wavelengths down to the visible range, due to their ultrathin nature achieving high-performance optical elements with good phase agility is still questionable \cite{van2020exciton}.\\
More recently, metasurfaces have shown great potentials for optical pulse shaping applications by finely tailoring the temporal profile of a near-infrared femtosecond pulse \cite{divitt2019ultrafast}. In analogy to the spatial analog computing approaches, such compelling architectures could be exploited to realize temporal analog signal processing including differentiation, integration, and Hilbert transformation which have been already demonstrated in photonic integrated circuit platforms \cite{liu2016fully,slavik2008photonic}. In this regard, leveraging powerful, elegant techniques such as dispersive Fourier transformation governing dispersive medium with high group-delay dispersion benefiting from time lens for chirp modulation would be useful \cite{goda2013dispersive,foster2008silicon}. \\
Optical neural networks based on subwavelength diffractive optics have gained significant attentions lately. Majumdar and his colleagues proposed an optical frontend for a convolutional neural network to alleviate the substantial challenges with the energy consumption and latency of existing electronic-based networks \cite{colburn2019optical}. They leveraged meta-optics to implement a dense architecture of 4\textit{f} corrolator that incorporates a general complex-valued transparent mask which is physical equivalent of the Fourier transform of the desired kernel. The optical frontend, that carries the predominant burden of computation, is coupled to software-based implementations of subsequent layers (including nonlinearity) to compete with the fully electronic networks dealing with large date sets. A useful comparison assessment on the classification accuracy of MNIST data as well as insightful discussions on the fundamental information capacity of the system are also provided. The introduced platform can be leveraged for practical implementation of hybrid photonic-electronic neural networks \cite{sui2020review,moughames2020three,zhang2019artificial}.\\
Extending the applicability of mathematical operations to the microwave wavelengths could take advantage of the ease of fabrication and simpler characterization of artificial structures \cite{ding2020metasurface}. To showcase the effectiveness of quantum analog behaviors in wave-based signal processing, microwave metamaterials were utilized as quantum searching simulators \cite{zhang2018implementing}. The explored dielectric structure, that relies on a 3D-printing technology, performs Grover's search algorithm, as a fast solver paradigm with quadratic speedup over classical methods and comparable to well-known quantum computing. The hardware implementation of other quantum algorithms such as Deutsch–Jozsa algorithm has been also experimentally demonstrated more recently \cite{cheng2020simulate}. Such compelling implementations reveal the potential of computational meta-structures in realization of beyond-classical mathematical operations. In a distinct work, using a single programmable digital metasurface architecture with spatio-temporal response, the authors in Ref. \cite{rajabalipanah2020space} realized multiple analog signal processing functions at microwave frequencies with specific application to detect the edges for sharp changes in the incident field.

Beyond the optical and microwave analog computing, recently, acoustic signal shows strong potential to enable complex computing applications and signal processing. Leveraging crystal of interconnecting pipes \cite{zangeneh2018performing}, topological insulators \cite{zangeneh2019topological}, layered labyrinthine metamaterials \cite{zuo2017mathematical,zuo2018acoustic,zuo2018acoustic}, people have shown several interesting results in this promising field.

Historically, most existing optical computing works are in the analog, continuous domain. However, there are some significant drawbacks to optical analog processing including limited flexibility, noise susceptibility, approximation error, and input-output device limitations \cite{sawchuk1984digital}. Although digital optical processors have been firstly introduced to circumvent the first three drawbacks to analog systems, there are still several fundamental physical constraints (imposed by quantum mechanics and thermodynamics) and practical limitations (such as lengths of communication lines and the corresponding signal transit times) with them.

\noindent \textbf{Acknowledgment:} 
This work was  supported by Office  of  Naval  Research (ONR)(N00014-18-1-2055, Dr. B. Bennett).

\noindent \textbf{Conflict of interest:} 
The authors declare no competing financial interest.



\providecommand{\noopsort}[1]{}\providecommand{\singleletter}[1]{#1}

\end{document}